\documentclass{iopart}
\usepackage{iopams}
\usepackage{graphicx}
\usepackage[tight]{subfigure}
\usepackage{textcomp}
\usepackage[normalem]{ulem}
\usepackage{hyperref}

\begin{document}

\title{Hyperfine local probe study of alkaline-earth 
manganites SrMnO$_3$ and BaMnO$_3$}

\author{J. N. Gon\c{c}alves, V. S. Amaral}
\address{Departamento de F\'isica and CICECO, Universidade de Aveiro,
3810-193 Aveiro, Portugal}
\ead{joaonsg@gmail.com}

\author{J. G. Correia}
\address{Centro de Ci\^encias e Tecnologias Nucleares, Instituto Superior T\'ecnico, Universidade de Lisboa, Estrada Nacional 10, km 139,7 2695-066 Bobadela LRS, Portugal}

\author{A. M. L. Lopes}
\address{Centro de F\'isica Nuclear da Universidade de Lisboa, 1649-003
Lisboa, Portugal}

\author{J. P. Ara\'ujo}
\address{IFIMUP and IN-Institute of Nanoscience and Nanotechnology, 
Departamento de F\'isica e Astronomia da FCUP,
Universidade do Porto, 4169-007 Porto, Portugal}

\author{P. B. Tavares}
\address{Departamento de Qu\'imica and CQ-VR, Universidade de
Tr\'as-os-Montes e Alto Douro, 1013, 5001-801 Vila Real, Portugal}


\begin{abstract}
We report perturbed angular correlation measurements with $^{111m}$Cd/$^{111}$Cd
and $^{111}$In/$^{111}$Cd probes, 
at the ISOLDE-CERN facility, 
in the manganite compounds BaMnO$_3$, with the 6H and 15R polymorphs, and
SrMnO$_3$, with the 4H polymorph.  
The electric field gradient (EFG) is measured, 
and found approximately constant in a large temperature range for all the
compounds. 
The EFG is also calculated from first-principles 
with density functional theory, 
and compared with the experimental results, 
by considering diluted substitutional Cd impurities. 
Based on the results, we assign as sites for the probes
the Ba (for BaMnO$_3$-6H, 15R), and Sr (for SrMnO$_3$-4H) sites, 
apart from fractions of undetermined origin in the case of BaMnO$_3$-6H. 
We predict the hyperfine parameters in the recently synthesized multiferroic 
manganite Sr$_{0.5}$Ba$_{0.5}$MnO$_{3}$, and its 
variation with the structure and electric polarization, 
which is found to be very small.
\end{abstract}

\pacs{31.30.Gs, 75.47.Lx, 71.15.Mb}

\maketitle


\section{Introduction}
Manganites have been the subject of renewed interest in the last years, 
due to the coupling of various order parameters, 
resulting in effects such as colossal magnetoresistance and multiferroicity. 
The divalent alkaline-earth based manganites, 
of the form AMnO$_3$ (where A is Ca, Sr or Ba) 
have a simpler electronic structure than the rare-earth based manganites, 
since the Mn ions can be considered formally as Mn$^{4+}$, 
and there  is no Jahn-Teller distortion resulting from partially filled $d$
orbitals. 
In these compounds, the observed structure is related to the ion sizes. 
The relation between interatomic distances and structure can be understood
by the tolerance factor
$t=d_{A\_O}/(\sqrt{2}d_{Mn\_O})$~\cite{Goldschmidt1926}, 
where $d_{A\_O}$ and $d_{Mn\_O}$ are the interatomic distances: 
$t=1$ corresponds to the ideal perovskite structure; 
values greater than one are found in hexagonal manganites, 
where the number of MnO$_{6}$ octahedra which share faces instead of vertices
(as in the ideal perovskite) is increased; 
values of $t$ smaller than one are concomitant with an orthorhombic structure. 
In its ground state at low temperature, 
CaMnO$_3$ is orthorhombic, due to the smaller radius of Ca$^{2+}$ (and tolerance
factor), 
while SrMnO$_3$ and BaMnO$_3$ are hexagonal. 
In spite of this apparent simplicity, 
the synthesis of the Sr or Ba systems is quite complex. 
An accurate control of the thermodynamic conditions, 
temperature and oxygen partial pressure, 
in the preparation and in the cooling steps (slow cool or quenching) is very
important to obtain a given phase. 
The Ba manganite, for example, can be synthesized in the non-stoichiometric form
BaMnO$_{3-x}$ 
with different values of $x$ (e.g.\ Adkin and Hayward~\cite{Adkin2007}).
  
\begin{figure}
\centering
\subfigure{
\includegraphics[width=0.20\linewidth]{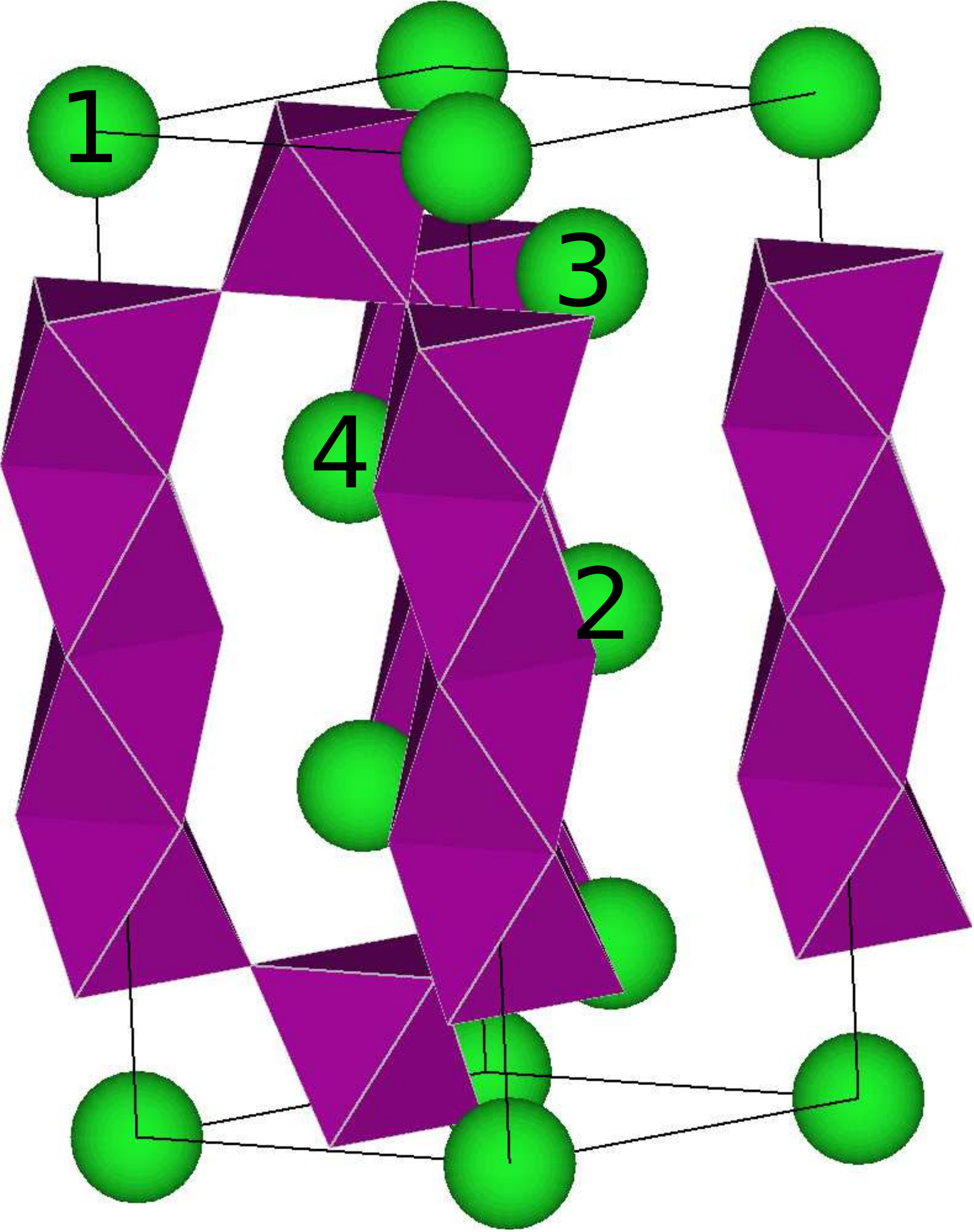}
}
\subfigure{
\includegraphics[width=0.20\linewidth]{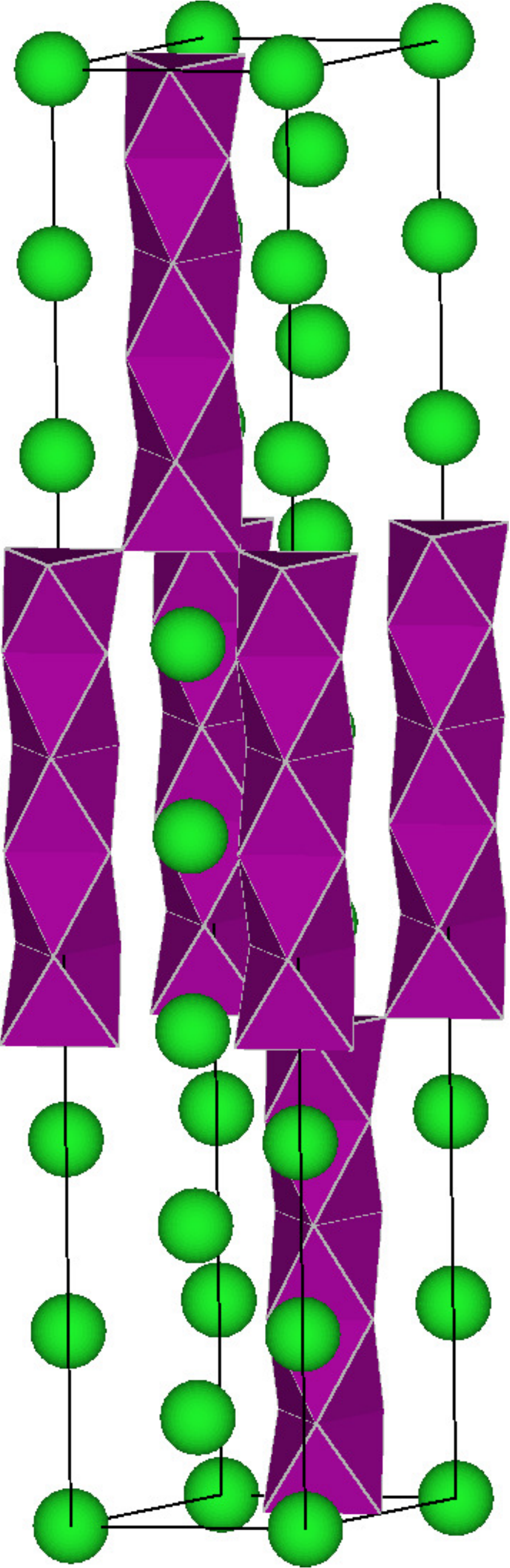}
}
\subfigure{
\includegraphics[width=0.20\linewidth]{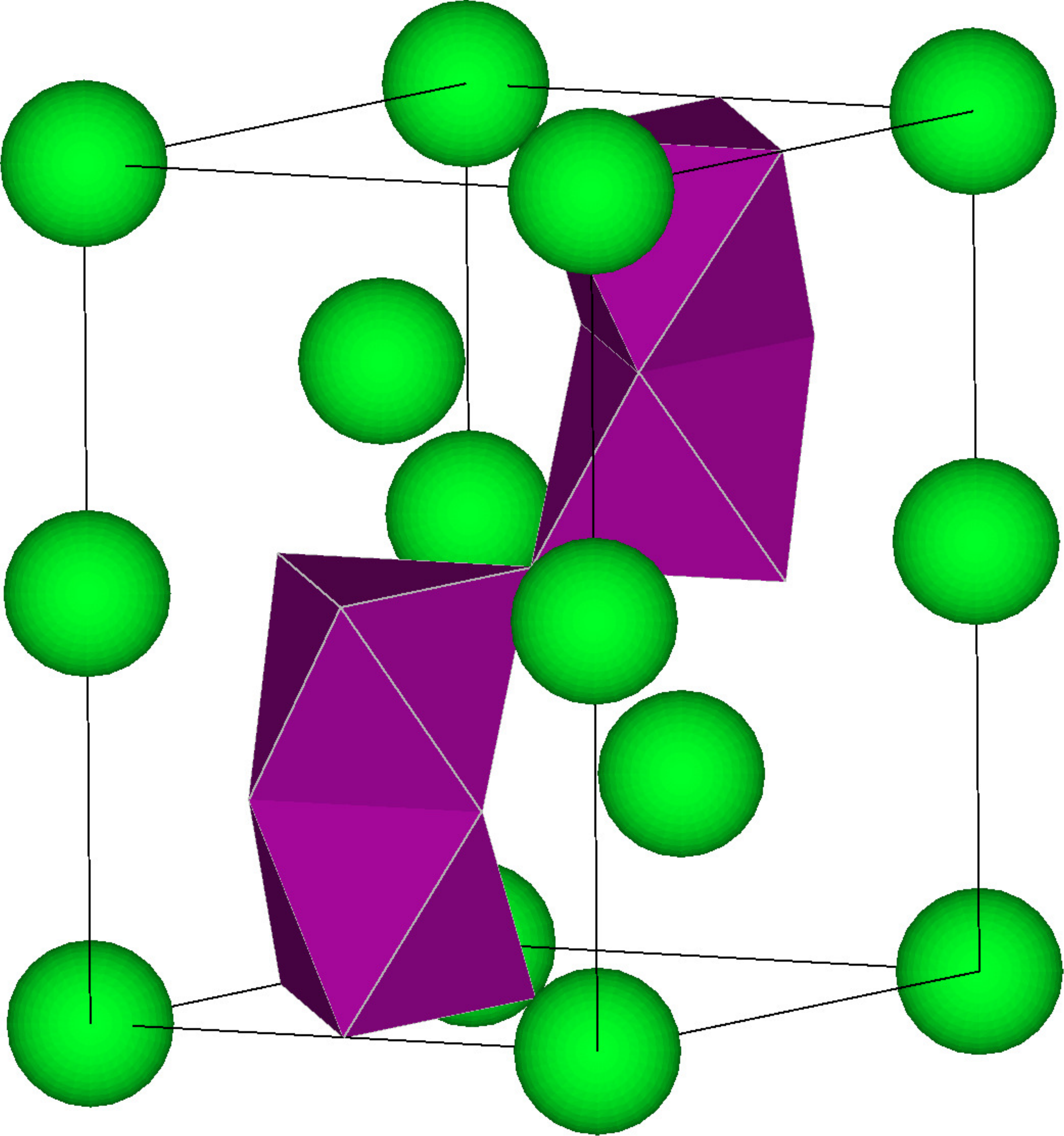}
}
\caption{Structural representation of the unit cells of
BaMnO$_3$-6H (left),
BaMnO$_3$-15R, in the hexagonal setting (center) 
and SrMnO$_3$-4H (right). The spheres are Ba/Sr
atoms, and the face- and corner-sharing octahedra are Mn centered and with O
atoms at their vertices.}
\label{structs}
\end{figure}

Various polymorphs of hexagonal structures can be synthesized, 
corresponding to different sizes for one-dimensional strings of face-sharing
octahedra, for example: 
hexagonal two-layer 2H (AB), where strings of face-sharing Mn-O octahedra have
infinite length, 
or hexagonal 6H (ABCACB), hexagonal 4H (ABAC), and rhombohedral 15R,
 where these strings are periodically broken (see figure~\ref{structs}).

There are some experimental measurements of structural and magnetic
properties carried on different forms of
BaMnO$_3$~\cite{Cussen2000,Chamberland1970,Hardy1962,Negas1971,Christensen1972}
and SrMnO$_3$.~\cite{Chamberland1970,Battle1988} The computational methods that
enable first-principles calculations have also given a good understanding of
these compounds. Density functional theory (DFT) calculations have been
performed for the study of these systems, with detailed analysis of the density
of states, band gaps, band structures and chemical bonding in the Ca/Ba/Sr
manganites~\cite{Sondena2007}, of the magnetic and structural
properties~\cite{Sondena2006}, and heat capacity and lattice
dynamics~\cite{Sondena2007_2} of cubic and hexagonal SrMnO$_3$. Hartree-Fock
calculations have also been used for the study of magnetic and electronic
properties in CaMnO$_3$~\cite{Fava1997}.
These works showed good reliability of the calculations.

There have been interesting developments in the perovskite 
type structure form of these compounds:  
recently, DFT studies have predicted ferroelectric instabilities in specific
strain conditions for CaMnO$_3$~\cite{Bhattacharjee2009},
SrMnO$_3$~\cite{Lee2010}, and
BaMnO$_3$~\cite{Rondinelli2009}, giving these  materials 
the possibility of being multiferroics. More recently, this promise was realized 
with the perovskite mixed manganite Sr$_{1/2}$Ba$_{1/2}$MnO$_3$, 
which is ferroelectric, with large
polarization due to displacement-type Mn off-centering, and possesses large
magnetoelectric effects~\cite{Sakai2011}. A DFT and model 
analysis~\cite{Giovannetti2012} traced 
the ferroelectricity to the Mn-O in-plane hopping which is 
distortion dependent, and has shown 
that ferroelectricity 
appears in the magnetic ordered phase 
despite a large negative magnetoelectric coupling.

In this work, we will use a combined theoretical (DFT calculations) 
and experimental hyperfine [perturbed angular correlation 
spectroscopy (PAC)] approach. 
In a previous work~\cite{Gonçalves2010}, some of us have presented
first-principles calculations on the
CaMnO$_3$ system, with the goal of understanding PAC results with a
$^{111m}$Cd$\rightarrow ^{111}$Cd probe. This probe has the particular advantage
of keeping
the element unchanged during the measurement. We considered the incorporation of
a Cd impurity in the manganite matrix, to calculate the properties at the probe
location. Here we use the same approach for the
first-principles calculations, and expand the PAC
study to the other alkaline-earth manganites, of Ba and Sr, with the same
$^{111m}$Cd$\rightarrow^{111}$Cd probe, and also with the
$^{111}$In$\rightarrow^{111}$Cd probe in
one case. Monophasic samples were synthesized: BaMnO$_3$ in 
the 6H phase and 15R phases,
and SrMnO$_3$ in the 4H phase.
We will direct our first-principles 
calculations to the electric field gradient
(EFG), which is compared directly to the experimental PAC data. 
We selected these three single phase samples in order to have 
examples of different polymorphs, to see to what 
extent the different kinds of local environments affect the EFG, allowing
us to simplify the fits and analysis of the spectra in each case.
The diagonalized EFG tensor, with components conventionally named according to
their magnitudes 
$|V_{zz}| \ge |V_{yy}| \ge |V_{xx}|$, is usually characterized by $V_{zz}$ and
the asymmetry parameter
$\eta=(V_{xx}-V_{yy})/V_{zz}$.
A local hyperfine property has several advantages, e.g., it can be related to
chemical
bonding and structure and can distinguish between different local environments, 
such as different lattice sites, and local defects in the probe's
neighborhood. Since the EFG depends sensitively on the electron
density its calculation will also be a good test of the theory in these systems.

Finally, due to the interest in the magnetoelectric 
coupling in this family of materials, 
the above mentioned multiferroic will 
be studied by \textit{ab initio} calculations, in order 
to predict the measured EFGs in this compound, and its variation 
with the structural degrees of freedom.

In the next sections we present results of calculations for SrMnO$_3$
(section~\ref{fpcalcSMO}) and for
different polymorphs of BaMnO$_3$ (section~\ref{fpcalcBMO}), and discuss the
results.
Direct comparison with the experiments is done by performing the calculations
with diluted Cd supercells, for the cases of 4H-SrMnO$_3$ and 6H-BaMnO$_3$.
The experimental results are presented in section~\ref{experiments} in the same
order, with its interpretation in relation to calculations. 
In section~\ref{SBMO} we show our analysis of EFG calculations in the  
multiferroic perovskite manganite 
Sr$_{0.5}$Ba$_{0.5}$MnO$_3$~\cite{Sakai2011}, 
and its possible variation with the polarization and structural changes.
 We conclude with a summary.

\section{First-principles calculations - hexagonal manganites}\label{fpcalc}
For the study of hexagonal manganites, we use the 
mixed LAPW/APW+lo full potential method, as implemented in the
\textsc{wien2k} package~\cite{WIEN2k}, which 
is known to provide accurate results, including the
hyperfine parameters. 
In this method the space is divided in atom-centered 
spheres and the interstitial space. The
wave functions inside the spheres are expanded in atomic like functions, linear
combinations of 
spherical harmonics times radial functions. 
In the interstitial region the wave functions are expanded 
in plane waves. 
Several k-points grids and maximum wave numbers for the
plane waves were tested in the simple cases to obtain a good convergence of
total
energy and EFGs. When calculating larger supercells the number of k-points is
diminished  
in conformity with the increasing size of the supercell. 
The GGA functional from Wu and Cohen~\cite{Wu2006} is used for all the
calculations except if noted otherwise.
Core states are treated fully-relativistically while valence states are treated
in the scalar-relativistic approximation. 

The \textsc{vasp} code~\cite{Kresse1996} also 
allows calculation of EFG tensors, and it will 
be used later (section~\ref{SBMO}) on in the multiferroic 
compound. It has been shown that both 
codes produce similar EFG values variations 
with polarization in ferroelectric perovskites~\cite{Gonçalves2012}.


\subsection{SrMnO$_3$}\label{fpcalcSMO}

SrMnO$_3$ with the 4H structure has the space group number 194, $P6_3/mmc$. 
The Wyckoff positions 
and fractional coordinates of the atoms in this structure are
the following: Sr1: 2a; Sr2: 2c; Mn: 4f ($1/3$, $2/3$, $z_{\mathrm{Mn}}$); O1:
6g; O2: 6h
(-$x_{\mathrm{O}}$, $-2x_{\mathrm{O}}$, $1/4$). 
With the experimental data of Battle et
al.~\cite{Battle1988} $z_{\mathrm{Mn}}=0.6122(5)$ and
$x_{\mathrm{O}}=0.1807(2)$, and the lattice
parameters are $a = 5.4434(2)$ \AA{}, $c = 9.0704(1)$ \AA{}. We used this
structure as a starting point for our calculations. After relaxing the free
atomic coordinates to calculated atomic forces less than $0.1$ eV/\AA{}, the
atomic
coordinates are $z_{\mathrm{Mn}}=0.61341$ and $x_{\mathrm{O}}=0.18058$. This
corresponds to small
variations of $0.12\%$ and $0.01\%$ for $z_{\mathrm{Mn}}$ and $x_{\mathrm{O}}$.
The small variation
of $x_\mathrm{O}$ is within the experimental error. The lattice parameters were
not
optimized, since usually the experimental ones are more dependable. In
this case we found
that a mesh of $6\times 6\times 3$ k-points and 
$RK_{max} = 6.5$ (where $R$ is the smallest muffin-tin sphere radius, 
and $K_{max}$ is the largest wave number used in the plane wave expansion) 
is enough for reasonably converged results, but more 
stringent criteria are used in the actual results. 
Table~\ref{SMO4HEFGFAAF} shows that the EFG values are similar for both
experimental and theoretically
relaxed structures, when assuming ferromagnetic (F) arrangement of Mn atoms, as
should be expected if the calculations are accurate. (Units of
$10^{21}$ V\,m$^{-2}$ will be used throughout this paper.)

We also considered the magnetic state where neighbor Mn planes in
the $z$ direction have opposite spin polarization. 
This antiferromagnetism,
called A-AF~\cite{Wollan1955}, is the  stable magnetic order at low
temperatures. For this case the deviation of the theoretically optimized
structure in relation to the initial experimental structure is even smaller, as
expected. With the experimental structure, 
this order is stable in relation to the
ferromagnetic state by $0.19$ eV/f.u. according 
to our calculations. The structural
coordinates are not moved more than 0.09\% relative to the experimental value in
the case of $z_{\mathrm{Mn}}$. Both parameters are equal to the experimental
ones within
the experimental error. This translates into almost exactly equal hyperfine
parameters for the theoretical and experimental structures,
even more so than in the F case. Nevertheless, the agreement is already good for
the F state, therefore, in the following  calculations, when we introduce Cd
impurities,
ferromagnetism is considered for simplicity, and furthermore, because
in the temperature range where we will measure and compare with theory SrMnO$_3$
should be paramagnetic~\cite{Battle1988}.
We show that the influence of the different magnetic states in the EFG is small,
for all atoms, in table~\ref{SMO4HEFGFAAF}. 

\begin{table}[ht]\centering
\caption{Calculated $V_{zz}$($10^{21}$ V\,m$^{-2}$) of 4H-SrMnO$_3$, with the F
and 
A-AF magnetic configurations, experimental and relaxed structures. The asymmetry
$\eta$ is zero except for the oxygen atoms, where it is also shown in the right.
\label{SMO4HEFGFAAF}}
\begin{tabular}{lcccc}
\hline\hline
Atom &(F, exp.)&   (F, rel.)            & (A-AF,exp.) &(A-AF,rel.) \\
\hline
Sr 1       & 3.94        & 4.11       & 4.01       &  4.09        \\
Sr 2       & 1.63        & 1.39       & 1.90       &  1.79        \\
Mn         &-0.29        &-0.40       & -0.45      & -0.48        \\
O 1        & 7.66; 0.07  & 7.73; 0.06 & 7.56; 0.21 &  7.61; 0.21  \\
O 2        &-8.17; 0.01  &-8.10; 0.02 &-7.89; 0.10 & -7.84; 0.09  \\
Mn 2 (A-AF)&             &            &-0.45       & -0.47        \\
\end{tabular}
\end{table}

In order to reproduce the experimental situation of a highly diluted Cd probe
(approx.\ smaller than $10$ ppm),  
we start by assuming that 
the Cd occupies the Sr sites substitutionally. 
This substitution of Cd at the alkaline earth site
has been found in the Ca manganite~\cite{Gonçalves2010}, and is
usually found experimentally in other manganites as well~\cite{Lopes2008}.
It is also a good assumption based on the ionic picture
Sr$^{2+}$Mn$^{4+}$O$_3^{2-}$ 
and considering that Cd has ionic charge 2+.
Two cases are possible: Cd occupies the Sr1 site, or the Sr2 site (as defined
above).
\begin{figure}%
  \centering
\subfigure{\includegraphics[width=0.25\linewidth]{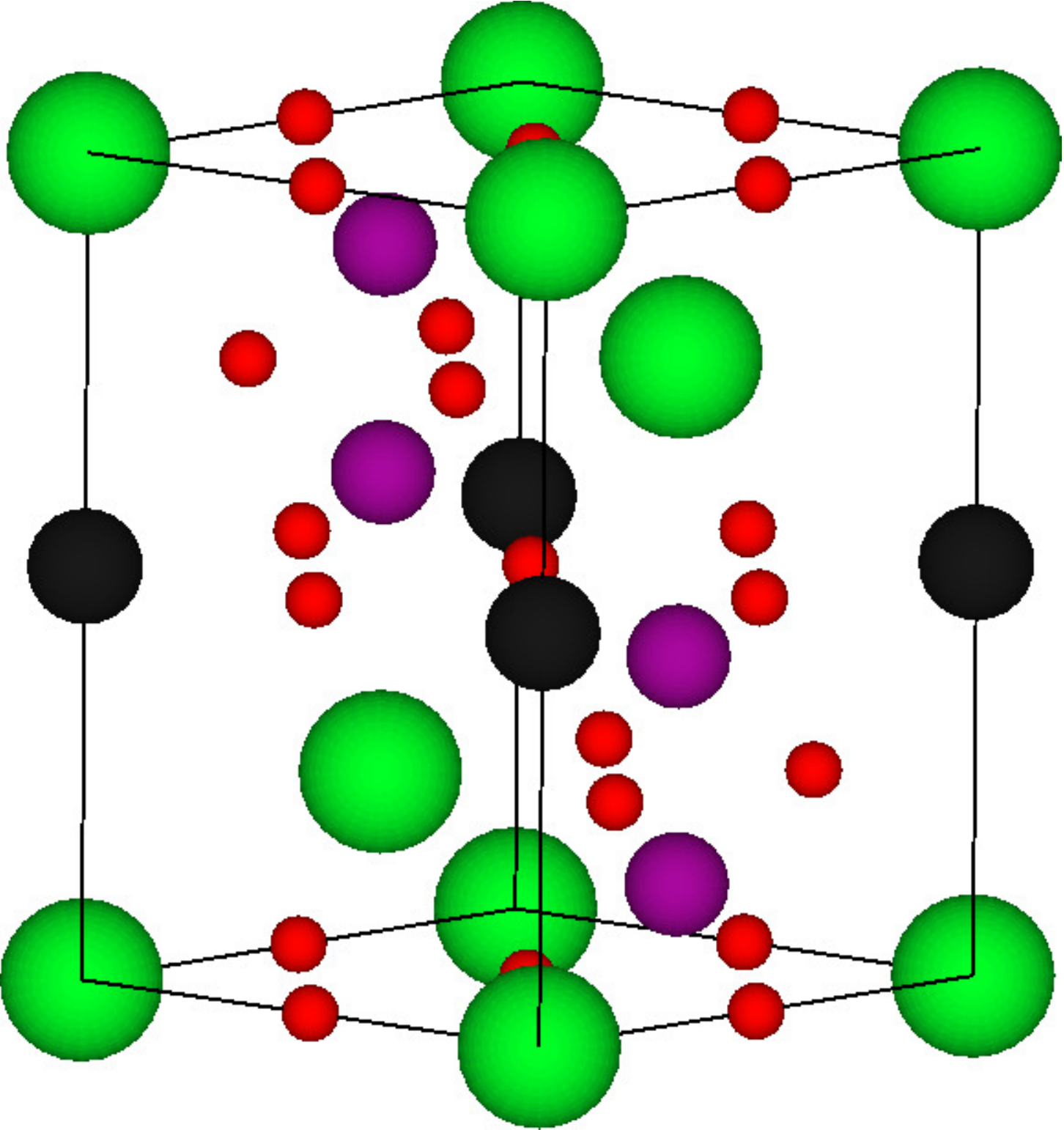}}
\subfigure{\includegraphics[width=0.25\linewidth]{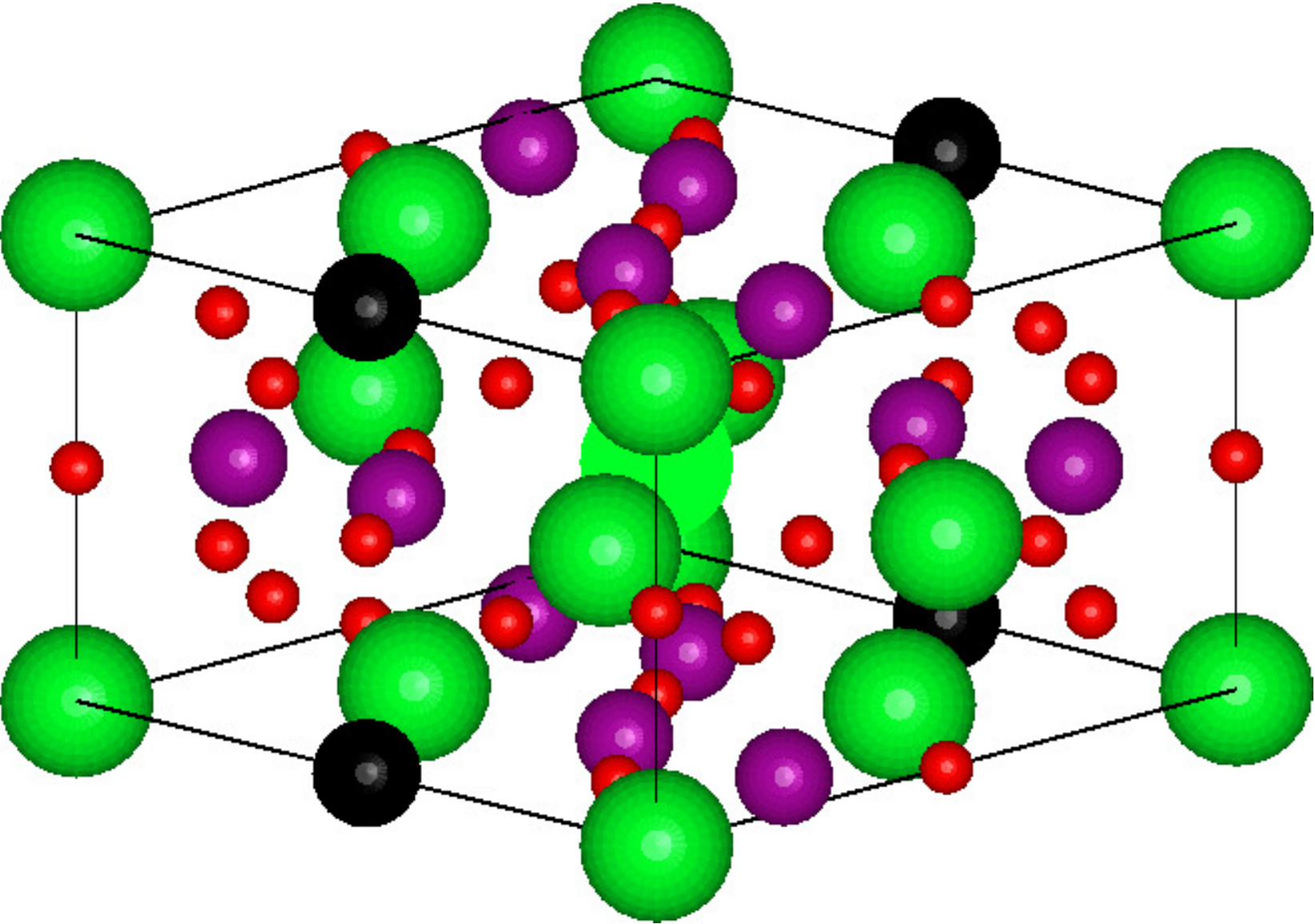}}
\subfigure{\includegraphics[width=0.25\linewidth]{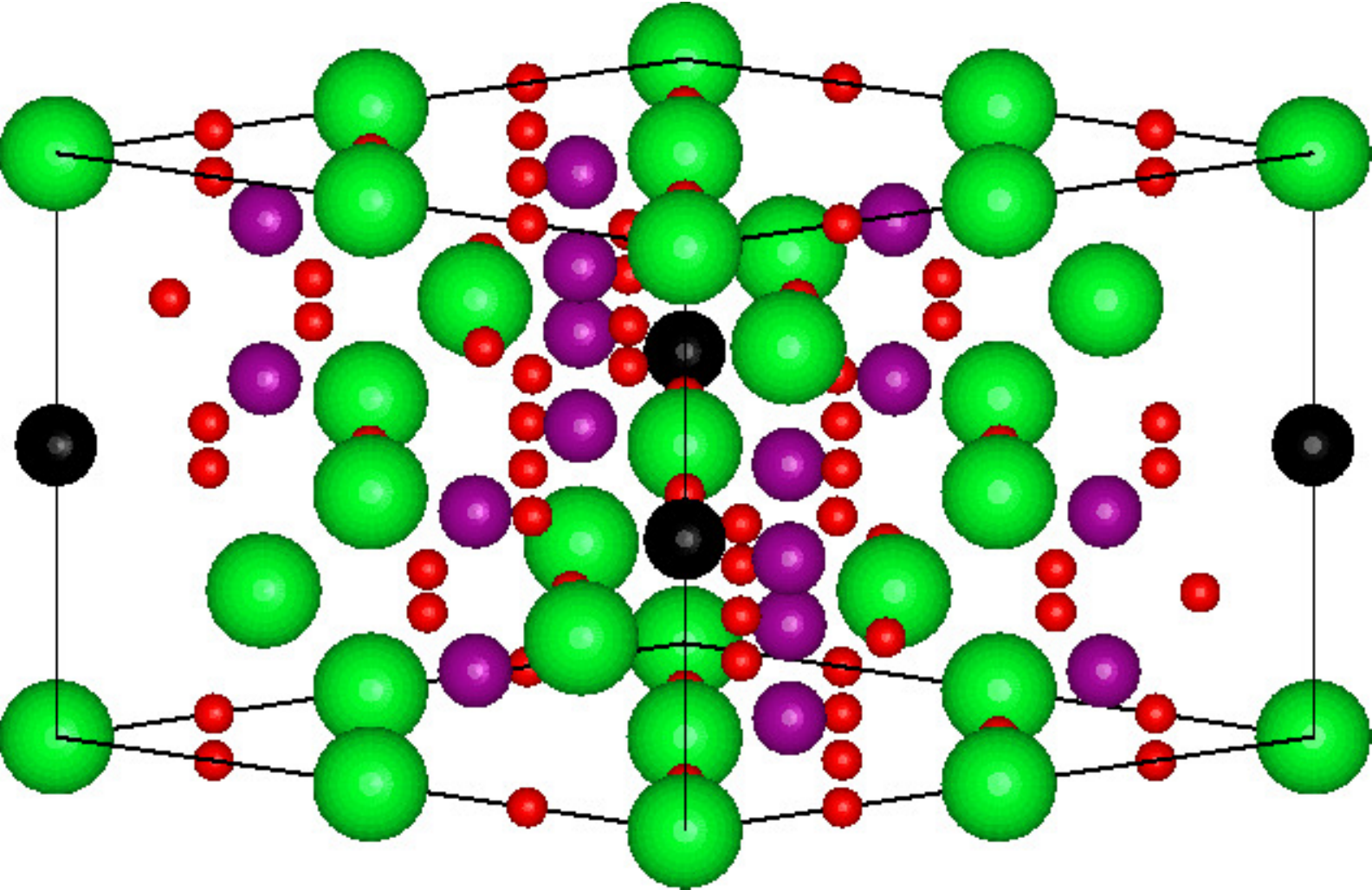}}
\caption{Supercells with Cd substitutional at the Sr 1 site. Cd
atoms are
represented by black spheres. Cd concentrations of 25\%, 12.5\% and 6.25\%.}
  \label{1sup}
\end{figure}
\begin{figure}
  \centering 
\subfigure{\includegraphics[width=0.25\linewidth]{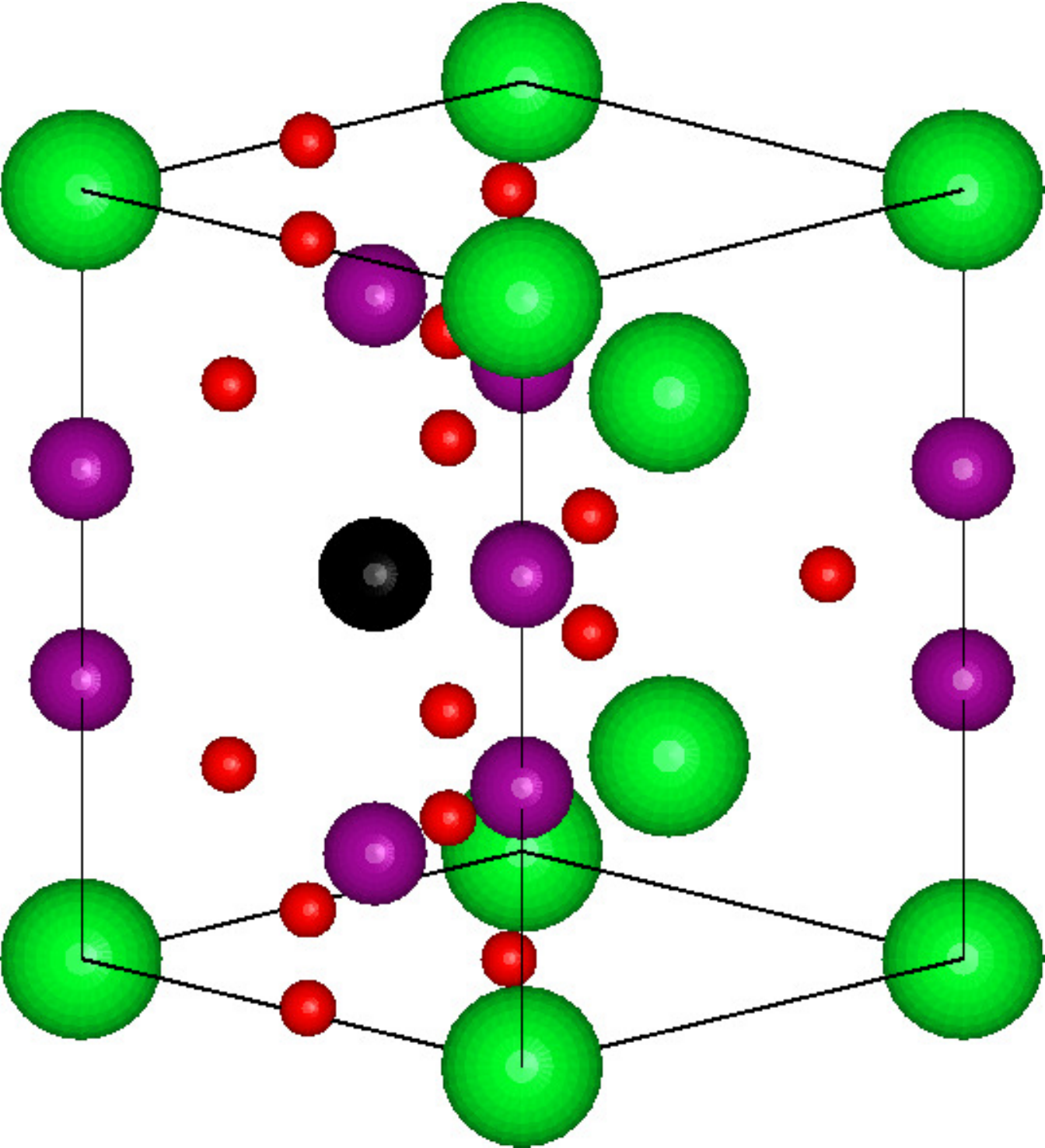}}
\subfigure{\includegraphics[width=0.25\linewidth]{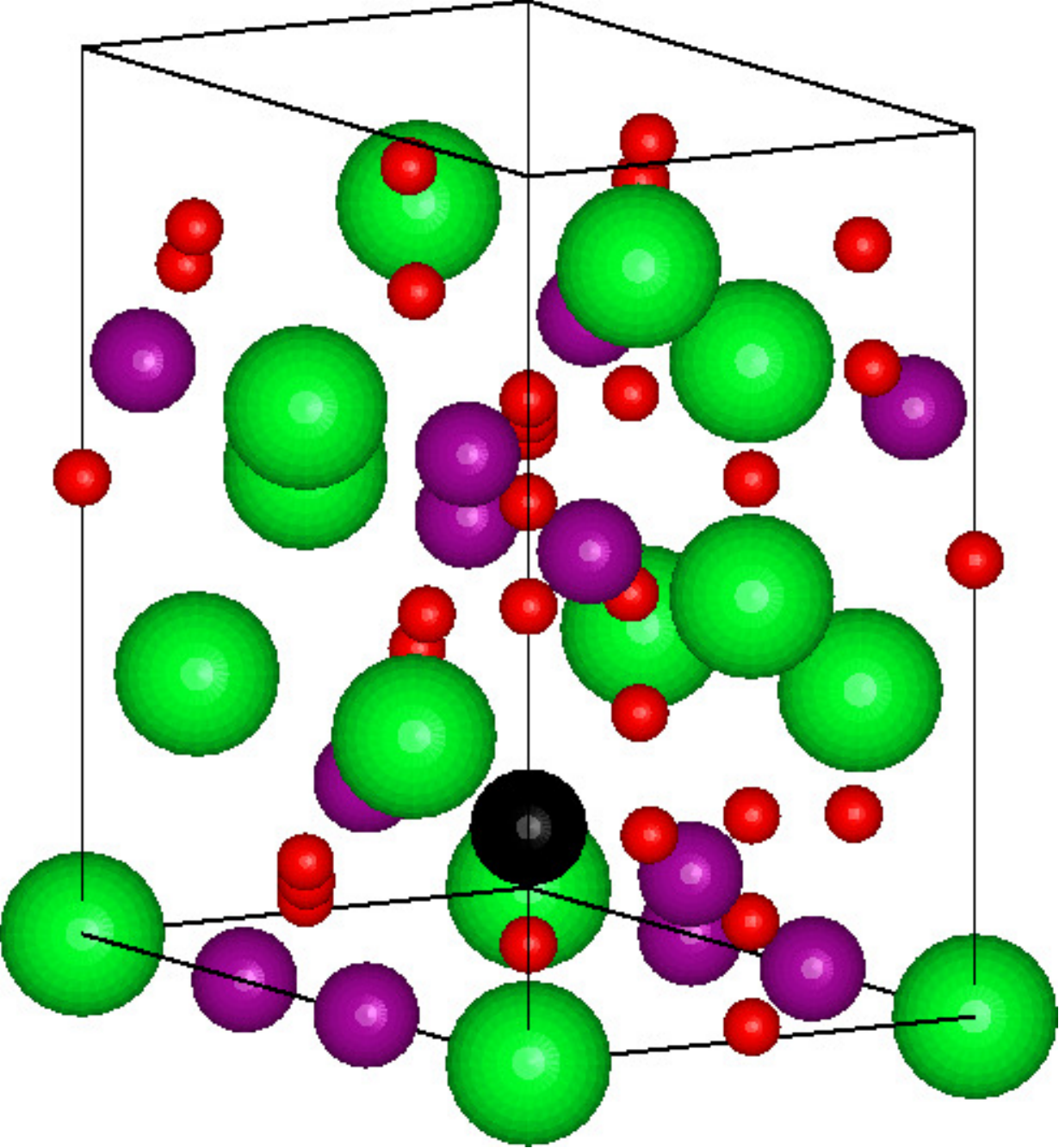}}
\caption{Supercells with Cd substitutional at the Sr 2 site. Cd
atoms are
represented by black spheres. Cd concentrations of 25\% and 12.5\%.}
\label{2sup}%
\end{figure}
These two situations are considered in independent calculations, 
and since dilutions of the order $10^{-6}$ are not feasible for calculations,
we use a series of increasingly larger supercells 
with one Cd to calculate different
impurity concentrations 
such that the EFG converges to the high dilution limit when Cd concentration is
lowered. (See Figs.~\ref{1sup} and \ref{2sup} for representations of some
supercells.)
As mentioned before, for this case we use only the F magnetic arrangement. 
With one atom of Cd substituting one of the sites in a single unit cell,
Sr$_{0.75}$Cd$_{0.25}$MnO$_3$ is obtained (25\% of Sr sites are substituted). If
we double the dimensions in the $x$ direction ($2\times1\times1$), we get
Sr$_{0.875}$Cd$_{0.125}$MnO$_3$, while doubling both $x$ and $y$ directions
gives a Sr$_{0.9375}$Cd$_{0.0625}$MnO$_3$ composition (6.25\%). Finally, for the
case of substitution at the Sr 1 site, we also doubled the previous supercell in
the $z$ direction, getting a $2\times2\times2$ 
supercell with 3.125\% concentration. The use
of high Cd concentrations when compared with experiment is justified by the
locality of the EFG, as the convergence of the results will show.

Table~\ref{SMO4HEFG} shows the calculated EFG for Cd at the Sr1 site, 
and the values of $V_{zz}$ appear to be approximately converged, around
$6\times10^{21}$ V\,m$^{-2}$ 
for concentrations of 12.5\% and lower. The calculation with higher Cd
dilution, 
with 3\% (1/32) of Cd atoms substituted at Sr sites, gives
$V_{zz}=6.22$.  
$\eta=0$ due to the hexagonal symmetry, 
except for the 2x1x1 supercell (where $a\neq b$) constructed for
Sr$_{0.875}$Cd$_{0.125}$MnO$_3$, where $\eta=0.34$. 
Table~\ref{SMO4HEFG} also contains the same type of calculations for the case
where Cd is substitutional at the Sr2 site. 
In this situation  $V_{zz}$ is fairly converged
at a value of the order of $1.5$ when 
considering the $2\times2\times1$ supercell
($x=0.125$). 

The calculated atomic forces are very low with just a maximum small force
of 0.26 eV/\AA{} in one of the sites. 
 Comparing the $V_{zz}$ values obtained with Cd and the $V_{zz}$ at the
corresponding Sr atom in the pure compound, 
we find that the value remains of the same order at the Sr2 site, 
but at the Sr1 site the value of $V_{zz}$ with Cd is considerably higher. 
The changes due to Cd substitution, although small, are different for each site,
showing that 
structural and electronic relaxations are essential to obtain accurate values.

The results for both cases with 6.25\% of Cd have the same energy, 
within the accuracy of the calculations, 
with a $3$ meV/f.u.\ difference.


\begin{table}[ht]\centering
\caption{\label{SMO4HEFG}Calculated $V_{zz}$ ($10^{21}$ V\,m$^{-2}$) of
4H-SrMnO$_3$, 
for a Cd atom substitutional at the Sr1 or Sr2 sites,
for various Cd concentrations $x$, in the formula Sr$_{1-x}$Cd$_x$MnO$_3$. }
\begin{tabular}{lrr}
\hline\hline
$x$         & $V_{zz}$  (Sr1) & $V_{zz}$  (Sr2)  \\
\hline
0.25        &         3.85      & 0.36         \\
0.125       &         6.08      & 1.38       \\
0.0625      &         6.47      & 1.48        \\
0.03125     &         6.22               \\
\end{tabular}
\end{table}

\subsection{BaMnO$_3$}\label{fpcalcBMO}

For the 6H-BaMnO$_3$ compound, the starting point for 
the 6H structure was taken from the work of Adkin et
al.~\cite{Adkin2007}. This structure has the $P\overline{6}m2$ space group, 
lattice parameters $a=5.6623(1)$, $c=13.9993(3) $\AA{}. 
The Wyckoff positions are: 
Ba1: 1a; Ba2: 1d; Ba3: 2h ($1/3$, $2/3$, $z_{\mathrm{Ba3}}$); 
Ba4: 2i ($2/3$, $1/3$, $z_{\mathrm{Ba4}}$); 
Mn1: 2g ($0$, $0$, $z_{\mathrm{Mn1}}$); Mn2: 
2g ($0$, $0$, $z_{\mathrm{Mn2}}$); Mn3: 2i ($2/3$, $1/3$, $z_{\mathrm{Mn3}}$); 
O1: 3j ($x_{\mathrm{O1}}$, $-x_{\mathrm{O1}}$, 0); 
O2: 3k ($x_{\mathrm{O2}}$, $-x_{\mathrm{O2}}$, $1/2$); 
O3: 6n ($x_{\mathrm{O3}}$, $-x_{\mathrm{O3}}$, z$_{\mathrm{O3}}$); 
O4: 6n ($x_{\mathrm{O4}}$, $-x_{\mathrm{O4}}$, z$_{\mathrm{O4}}$). 
The experimental data is 
$z_{\mathrm{Ba3}}=0.1667(2)$, $z_{\mathrm{Ba4}}=0.3365(1)$, 
$z_{\mathrm{Mn1}}=0.2364(1)$, $z_{\mathrm{Mn2}}=0.4135(2)$, 
$z_{\mathrm{Mn3}}=0.0906(1)$,
$x_{\mathrm{O1}}=0.5215(1)$, $x_{\mathrm{O2}}=0.8503(1)$, 
$x_{\mathrm{O3}}=0.8830(1)$, 
$z_{\mathrm{O3}}=0.16847(9)$, $x_{\mathrm{O4}}=0.1486(1)$, 
$z_{\mathrm{O4}}=0.3273(4)$ 
(as displayed in table 4 of~\cite{Adkin2007}, 
however, note that the $y$ coordinates of three oxygen atoms should be negative).
Ferromagnetism was assumed, which is not a known state of BaMnO$_3$, 
but this should be reasonable judging by the the similar
system SrMnO$_3$-4H (shown previously) 
where the EFG has small changes with a different magnetic state. 
 20 k-points in the irreducible Brillouin zone
(but 5 k-points are already enough for converged results)
and a maximum wave number with $RK_{max}=7.5$ were used
($RK_{max}=7$, used in some supercells, already gives converged results).
With this calculation the atomic forces are generally low (one force
at
0.82 and the rest less than 0.33 eV/\AA{}).
The symmetrically unconstrained atomic parameters were permitted to vary
and a theoretically more stable structure was obtained,
where the atomic forces are lower than 0.05 eV/\AA{}.  The atomic coordinates
obtained have at most a 0.88\% difference with respect to the initial
experimental ones. 

Results of the EFG for the structure of Adkin et al. and after relaxation of
internal parameters are summarized in
Table~\ref{BMO6HEFG}.
We show also the results with the
GGA-PBE functional~\cite{Perdew1996}. There are almost no changes in the results
with the PBE, typically of the order or less than $0.1\times10^{21}$
V\,m$^{-2}$.
The asymmetry parameter is also shown for the sites where symmetry allows
it to be different from zero, and it is usually very small, except for one of
the oxygen atoms. 
We also performed a test including spin-orbit coupling, 
using the previously relaxed structure, but this also produced insignificant
changes in the results. 
Results for the EFG of the 15R phase and the hexagonal polymorphs 2H (which is
the stable phase at low temperatures) and 4H are shown in tables~\ref{BMO15REFG}
and~\ref{BMOpoliEFG}. The 2H structure is taken from
~\cite{Cussen2000} [space group $P63/mmc$, 
$a=5.6991(2)$, $c=4.8148(2)$\AA{}, 
with Wyckoff positions: Ba: 2d; Mn: 2a; O: 
6h ($x_\mathrm{O}$, 2$x_\mathrm{O}$, $1/4$), 
with $x_\mathrm{O}=0.14960(5)$] and the
4H structure is taken from our own X-ray diffraction 
measurements [space group $P63/mmc$, $a=5.6833$, $c=9.3556$\AA{}, 
with Wyckoff positions: Ba1: 2a; Ba2: 2c; 
Mn: 4f ($1/3$, $2/3$, $z_\mathrm{Mn}$); O: 6g, 
with $z_{\mathrm{Mn}}=0.6086$. 
$z_\mathrm{Mn}=0.6131$ after DFT relaxation]. The 15R
experimental structure is from Adkin et al.~\cite{Adkin2007} 
[space group $R\overline{3}m$, Wyckoff positions: Ba: 1a; Ba2: 
2c ($x_{\mathrm{Ba2}}$, $x_{\mathrm{Ba2}}$, 
$x_{\mathrm{Ba2}}$); Ba3: 2c ($x_{\mathrm{Ba3}}$, 
$x_{\mathrm{Ba3}}$, $x_{\mathrm{Ba3}}$); 
Mn1: 1b; Mn2: 2c ($x_{\mathrm{Mn2}}$, $x_{\mathrm{Mn2}}$, 
$x_{\mathrm{Mn2}}$); Mn3: ($x_{\mathrm{Mn3}}$, 
$x_{\mathrm{Mn3}}$, $x_{\mathrm{Mn3}}$); 
O1: 3e; O2: 6h ($x_{\mathrm{O2}}$, $x_{\mathrm{O2}}$, 
$z_{\mathrm{O2}}$); O3: 6h ($x_{\mathrm{O3}}$, 
$x_{\mathrm{O3}}$, $z_{\mathrm{O3}}$), 
with the experimental values being 
$x_{\mathrm{Ba2}}=0.1337$, $x_{\mathrm{Ba3}}=0.2653$, $x_{\mathrm{Mn2}}=0.3618$, 
$x_{\mathrm{Mn3}}=0.4316$, $x_{\mathrm{O2}}=0.2502$, 
$z_{\mathrm{O2}}=0.69211$, $x_{\mathrm{O3}}=0.6156$, $z_{\mathrm{O3}}=0.1651$ 
(Table 2 from~\cite{Adkin2007} 
with coordinates in the hexagonal system). The atomic
coordinates were also optimized,
 and remain qualitatively equal to the experimental structure calculations. 

These calculations show that the EFG has variations
depending on the polymorph, particularly in the different atoms of Ba,  
due to the different packings of Mn-O octahedra. 
Mn atoms have $V_{zz}$ between -2.6 and 0.2 $\times 10^{21} $V\,m$^{-2}$ for all
polymorphs,
and O atoms have EFGs approximately between -9 and 9.5, and in absolute value 
are located in the range from 5.9 to 9.5.
 Ba has usually one high value between 10 and 13,
and for other Ba atoms the EFG can be quite low, depending on the structure.
\begin{table*}[ht]\centering
\caption{\label{BMO6HEFG}Calculated EFG of BaMnO$_3$, 6H
polymorph, experimental and relaxed structures, 
also with spin-orbit coupling included. 
When there are two sets of values, the second are obtained with the GGA-PBE
exchange-correlation functional ($V_{zz}$ in units of $10^{21}$ V\,m$^{-2}$).}
\begin{tabular}{lrrrrrr}
\hline\hline
Atom                |& $V_{zz}$ (exp) |& $\eta$ (exp.) |& $V_{zz}$ (rel) |&
$\eta$ (rel)  |& $V_{zz}$ (rel + so) |& $\eta$ (rel + so)\\
\hline
Ba 1 (\textit{chc}) |&  -0.26; -0.27  &   &    -0.22; -0.13       &            &
-0.24 & \\
Ba 2 (\textit{hhh}) |&  11.86; 11.82  &   &    13.05; 12.73       &            &
13.05 & \\
Ba 3 (\textit{hch}) |&   9.38; 9.38   &   &     9.09;  8.83       &            & 
9.10 & \\
Ba 4 (\textit{hhc})|&   5.01 ; 4.95  &    &     4.76;  4.67       &            & 
4.76 & \\
Mn 1 |&   0.03 ; -0.03 &              &     0.18;  0.10       &            & 
0.17 & \\
Mn 2 |&  -1.79; -1.91  &              &    -1.75; -1.85       &            &
-1.76 & \\
Mn 3 |&  -0.20; -0.27  &              &    -0.24; -0.31       &            &
-0.25 & \\
O  1 (\textit{chc}) |& -9.01; -9.12 & 0.02; 0.01  & -8.96; -9.08  & 0.01; 0.02 &
-8.95 & 0.01 \\
O  2 (\textit{hhh}) |& -6.01; -6.10 & 0.07; 0.09  & -6.05; -6.21  & 0.03; 0.02 &
-6.04 & 0.03 \\
O  3 (\textit{hch}) |& 8.49 ; 8.66  & 0.001; 0.005& 8.43;  8.69  & 0.003; 0.003& 
8.42 & 0.003 \\
O  4 (\textit{hhc}) |& -7.69 ; -7.78 & 0.38; 0.38 & -7.66; -7.80  & 0.36; 0.35 &
-7.65 & 0.36 \\
\hline\hline
\end{tabular}
\end{table*}

\begin{table}[ht]\centering
\caption{Calculated EFG of BaMnO$_3$, in the 15R polymorph, with 
relaxed atomic positions ($V_{zz}$ in $10^{21}$ V\,m$^{-2}$).}\label{BMO15REFG}
\begin{tabular}{lrr}
\hline\hline
Atom & $V_{zz}$ & $\eta$    \\
\hline
Ba 1(\textit{hch}) &               7.12           &             \\
Ba 2(\textit{hhh}) &              11.35           &             \\
Ba 3(\textit{hhc}) &              4.55           &             \\
Mn 1&             -1.73           &             \\
Mn 2 &              0.19           &             \\
Mn 3 &             -1.79           &             \\
O 1(\textit{hch}) &               8.43           &    0.06    \\
O 2(\textit{hhh}) &              -6.13           &    0.03    \\
O 3(\textit{hhc}) &              -7.79           &    0.35     \\
\end{tabular}
\end{table}
\begin{table}[ht]\centering
\caption{Calculated EFGs of BaMnO$_3$,
in the 4H (left), and 2H (right) structure from experimental and relaxed
structures. ($V_{zz}$ in $10^{21}$ V\,m$^{-2}$).}\label{BMOpoliEFG}
\begin{tabular}{lcccc}
\hline\hline
Atom  |&  $V_{zz}$ |& $\eta$ |& $V_{zz}$ |& $\eta$           \\
\hline 
Ba 1& 9.99; 11.93  & & 10.33; 10.22&                         \\
Ba 2&                          1.64; -0.38 &                 \\
Mn 1&  0.10; -0.29 &&-1.92; -1.90&                           \\
O  1& -9.45; -9.10 &0.08; 0.05    &-6.52; -6.58&0.04; 0.06   \\
O  2& 8.58; 8.56 &0.07; 0.06                                 \\
\end{tabular}
\end{table}

 With \textit{c} and \textit{h} denoting horizontal layers of Ba and O where two
MnO$_6$ octahedra share corners or faces, respectively, and  concerning the
different hexagonal polymorphs, there are 
four possibilities for local (3 layers) Ba/O environments, 
corresponding to the central layers of the following
sequences: \textit{chc}, \textit{hch}, \textit{hhc}, and \textit{hhh} (see
figure~8 of~\cite{Adkin2007}). In the 6H polymorph the four kinds of
environments interact, and correspond, in the same order, to the Ba1/O1, Ba3/O3,
Ba4/O4, and Ba2/O2 layers (see numbers in figure~\ref{structs}). 
The exact values for the same environment in the other polymorphs are
obviously different, due to the different interactions
between environments, however, there are general trends followed by the $V_{zz}$
of Ba atoms in the
different structures. In the 2H form, consisting of infinite chains of face
sharing octahedra along the
$z$ direction, there is only one type of Ba atom (\textit{hhh}), with the high
$V_{zz}$ value
reported (table~\ref{BMOpoliEFG}). This environment also
corresponds the highest value in BaMnO$_3$-6H. 
The second largest value in 6H polymorph corresponds to the Ba3 site, with a
\textit{hch} layer (which is a corner-sharing layer, but also surrounded by
face-sharing environments). In BaMnO$_3$-4H, where there
are \textit{hch} and \textit{chc} environments, \textit{hch} corresponds to
the largest value, with \textit{chc} corresponding to a smaller $V_{zz}$, which
also happens to be the smallest $V_{zz}$ in 6H. Finally, in the 15R structure
the following environments coexist: \textit{hhh}, \textit{hch}, and
\textit{hhc}. The same order is followed as in the 6H case, namely
$V_{zz}$(\textit{hhh})$ > V_{zz}$(\textit{hch})$ > V_{zz}$(\textit{hhc}). In
summary, considering the different Ba  environments, the following rule 
is general: $V_{zz}$(\textit{hhh})$ > V_{zz}$(\textit{hch})$ >
V{zz}$(\textit{hhc})$ > V_{zz}$(\textit{chc}): the more neighboring 
face-sharing octahedra, the higher $V_{zz}$ is. In the O environments,
although the values are closer, there is also a general rule for all the
polymorphs studied,  $V_{zz}$(\textit{hhh})$ < 
V_{zz}$(\textit{hhc})$ < V_{zz}$(\textit{hch})$ < V_{zz}$(\textit{chc}), 
considering only the absolute value. 
Considering the sign, it is notable that the $hch$ 
environment (in BaMnO$_3$-6H, BaMnO$_3$-15R, and 
BaMnO$_3$-4H) is a positive $V_{zz}$ while the other 
environments correspond to negative values. 
As for the Mn atoms, the ones centering 
corner-sharing octahedra have $V_{zz}\sim0$ while 
the others have $V_{zz}\sim-1.8$.
To finish this analysis, we can say this 
case clearly illustrates the locality of EFG, since 
sites with the same elements can be equated and distingued across polymorphs 
based only on their neighboring environments.

In order to compare with the 
PAC experiments with Cd probes,
we introduce Cd impurities substitutional at 
the Ba site. We consider the case of the BaMnO$_3$-6H structure, 
and note that the structure is already quite complicated without Cd, 
making large supercells computationally demanding. 
However, already with the unit cell we  can substitute 
one atom of Cd to get the concentration $x=1/6$. 
Having considered this fact we calculated the other Cd 
substitutions also for the high concentration $x=1/6$, 
in order to reduce computational time, expecting this to be a good approximation
to the low impurity concentration limit. 
We present the $V_{zz}$ obtained at the Cd site substituted 
in each of the inequivalent Ba sites in the table~\ref{BMO6HCd}. 
The asymmetry is zero at the Cd sites for all cases due to lattice symmetry.

The comparison of the EFG at Cd (table~\ref{BMO6HCd})
with the EFG at the
respective Ba sites in the pure compound (table~\ref{BMO6HEFG}) shows no
proportionality, meaning that, as expected due 
to the different local environments, 
the Cd impurity probes provoke different changes in structure 
and in the local charge density. We also show 
the results of the total energies for those 
four supercells, according to which the probe should 
occupy preferentially the Ba1 site, followed by the Ba3 site.

\begin{table}[ht]\centering
\caption{Calculated EFG of Ba$_{5/6}$Cd$_{1/6}$MnO$_3$, in the 6H polymorph, at
the Cd site, 
for independent calculations with Cd substituted at each of the inequivalent 6H
sites. Corresponding total energies, with respect to the highest.}
\label{BMO6HCd}
\begin{tabular}{lrr}
\hline\hline
Site with Cd &| $V_{zz}^{Cd}(10^{21}$ V\,m$^{-2})$ &| Total energy (eV) \\
\hline
Ba 1 &               -1.72        &    -0.65        \\
Ba 2 &                4.80        &     0           \\
Ba 3 &                2.27        &    -0.41        \\
Ba 4 &                0.20        &    -0.25        \\
\end{tabular}
\end{table}

These cells are complicated, and yet their size may not be enough 
for a good convergence of the EFG. To get a good value for the 
Cd at the (\textit{hhh}) environment, we used the simpler 2H polymorph
and constructed a $2\times2\times2$ supercell, where the Cd 
atoms are separated by approximately the same 
distances as in a $2\times2\times1$ 6H supercell. 
The result is $V_{zz}\sim8$, which 
shows that probably $V_{zz}^{Cd}$ at the Ba2 site would 
increase with the supercell size, 
and that the calculated values for Cd at BaMnO$_3$-6H can 
only be taken as a qualitative approximation.


\section{Experimental results - Hexagonal Manganites}\label{experiments}
Monophasic samples of  BaMnO$_3$ with
the 6H structure, other monophasic samples with the 15R structure, and
others of SrMnO$_3$ with the 4H structure were prepared with the urea sol-gel
combustion method~\cite{Ferreira2008} or with conventional solid state method. 

In the combustion method stoichiometric amounts of BaCO$_3$ and
Mn(NO$_3$)$_2\cdot$4H$_2$O (ABCR $> 98$\%) were 
dissolved in diluted nitric acid. The
pH of the solution was adjusted to 5.2 with ammonia solution. The added amount
of urea was calculated so that 3 moles of urea were presented for each mole of
cationic element (Ba+Mn). The solution was stirred and heated to evaporate all
water and decompose the urea. At the end of the process (when the temperature
reaches 200\textdegree C) the gel auto ignites and a controlled but fast
combustion (3 to 5 seconds) yielded a dark powder. This powder was calcinated at
600\textdegree C for 1 hour followed by 700\textdegree C for 5 minutes, grounded
with a mortar and a pestle, passed through a 38 $\mu m$ sieve and pelletized.
The pellet was thermal treated at 900\textdegree C for 40 hours regrounded with
a mortar and a pestle, repelletized and treated at 1400\textdegree C for 100
hours with quench cooling (less then 1 minute (yielding phase Ba-6H)).
For samples prepared by conventional solid state method, BaCO$_3$ (or SrCO$_3$)
and MnO$_2$ were mixed with ethanol and calcinated at 700\textdegree C during 40
hours. Thermal treatments were processed at 900\textdegree C, 1000\textdegree C
and 1100\textdegree C during 40 hours with intermediate grindings. Final
sintering was performed at 1275\textdegree C for 180 hours with quench cooling,
yielding phases Ba-15R and Sr-4H. 

The densities of the samples were measured by geometric factor using a vernier
calliper.
The x-ray diffraction patterns were obtained in $\theta$/2$\theta$ mode from 10
to 90\textdegree  in an X’Pert Pro diffractometer equipped with a X’Celerator
detector and a secondary monocromator. The data were analysed using the Powder
Cell 2.3 software~\cite{powdercell} and structural data from
Adkin~\cite{Adkin2007}. SEM was performed using a FEI Quanta 400 equipped with
an EDS from EDAX.

These samples were subsequently implanted with the radioactive $^{111m}$Cd and
$^{111}$In probes,
with concentrations in the order of ppm.
The isotope beams were produced at the ISOLDE-CERN facility, and the
implantation was done in high vacuum at room temperature, at 60 keV, with a dose
of approximately $1\times10^{12}$ atoms$\cdot$cm$^{-2}$.

In the $\gamma-\gamma$ time-differential perturbed angular 
correlation method, the directional
correlation of the two gamma decays in a decay cascade
of the radioisotope probe is perturbed by the hyperfine interactions, which
depend on the nuclear properties of
the isotope and on the extranuclear local environment~\cite{Schatz1996}. 
The $^{111m}$Cd probes decay to $^{111}$Cd with two consecutive $\gamma$ decays.
The intermediate state has a half life $t_{1/2}=45\ $ns while the parent state
decays with $t_{1/2}=48$ minutes. The quadrupole nuclear moment of the
intermediate nuclear state, relevant for the hyperfine interactions, is
$Q=+0.77(12)\ b$ (taken from~\cite{Stone2005}). The $^{111}$In probes
decay to the same intermediate state.

All of the relevant experimental information is condensed in  the  anisotropic
ratio function $R(t)$,  
expressed as $R(t)=\sum A_{kk}G_{kk}(t)$, a function of time, where $A_{kk}$
are called anisotropy coefficients and depend on the spins and multipolarity of
the $\gamma$ decays, and $G_{kk}(t)$ contains the information of the hyperfine
parameters,
relevant to the system under study. The anisotropy coefficients are shown in
table~\ref{anisotropy}. 

\begin{table}[ht]\centering
\caption{\label{anisotropy}Anisotropy coefficients of the $\gamma-\gamma$
cascade from $^{111m}$Cd$\rightarrow^{111}$Cd and
$^{111}$In$\rightarrow^{111}$Cd.}
\begin{tabular}{lrrrr}
\hline\hline
Parent nucleus & A$_{22}$ & A$_{24}$ & A$_{42}$ & A$_{44}$ \\
\hline
$^{111m}$Cd & 0.1258&-0.0956&-0.0006 &-0.0004\\
\hline
$^{111}$In &-0.1121& -0.0716& -0.0003&  -0.0004\\
\hline\hline
\end{tabular}
\end{table}

A numerical fit of the R(t) function with its Fourier analysis,
taking appropriate nuclear and transition parameters, yields the hyperfine
parameters, EFG and magnetic hyperfine field~\cite{Barradas1993}.
The quadrupole electric interaction, with additional knowledge of the probe's quadrupole moment, gives the EFG.

In our studies the magnetic hyperfine field is not detected.
SrMnO$_3$, known to have T$_N=278$~\cite{Battle1988} or $350$
K~\cite{Chamberland1970}, will be paramagnetic at all temperatures measured,
except for room temperature, where it may be antiferromagnetic.
For BaMnO$_3$, the N\'eel temperature depends of the polytype.
For the phases of interest, 6H and 15R, its values are of the order  250
K~\cite{Adkin2007}, so 
that also on almost all temperatures measured it is paramagnetic. Therefore we assume there are no magnetic interactions, and start by considering one EFG (environment), increasing the number of environments until no more are needed to account for the experimental spectrum. The frequency, asymmetry parameter, and shape of the distribution are fitted to the spectrum in different runs of the fit program, after which we estimate its error by simple statistics of the output values obtained when running many  independent fits starting from different, and reasonable, values of these parameters. 

Each measurement at different temperatures was done with a different sample,
this was required to get
good statistics in the lifetime of the isotope.
Preceding each measurement the samples were annealed in vacuum at 700\textdegree C during a
period of 20 minutes,
to remove defects from implantation. These conditions were found to be enough in
previous studies of other
manganites~\cite{Lopes2008,Lopes2006,Araujo2005,Araujo2001}.

\subsection{X-ray powder diffraction}

X-ray powder diffraction was used not only for the structural identification and
quality check of the samples, 
but also  to check that the conditions of implantation,
 annealing and measurement had no influence on the initial state of the samples.
The Cu anode was used, with wavelenghts $K\alpha_1=1.540598 $\AA{},
$K\alpha_2=1.544426 $\AA{}, and $K\beta=1.39225 $\AA{}. 
The scanning is continuous
with a step size of 0.001 degrees, and $15<2\theta<90$.
X-ray spectra were taken at various high temperatures, including several hours
at a temperature of 900\textdegree C, to
check that the annealing of 700\textdegree C,
during 20 minutes before the PAC\ measurement has no influence on the structure.
Refinement parameters of 
representative powder diffraction 
patterns are shown in table~\ref{XRD}.

 For BaMnO$_3$, the several possibilities of different phases
(4H, 7H, 15R, 6H, ...) 
that could appear are not easily distinguishable 
in the diffraction patterns so it is possible 
that some structural changes occurred with the implantation, annealing, and PAC
measurement. 
However, this is unlikely, given the consistent PAC results 
on the several temperatures measured for BaMnO$_3$-6H, as we will show. 
BaMnO$_3$-2H, the most stable low temperature structure, is easily
eliminated from this set of 
possible structures on all diffraction patterns.

\begin{table}[ht]\centering
\caption{Main refinement parameters of XRD for some of some of the studied
samples, and lattice parameters $a$ and $c$ (\AA). SrMnO$_3$-1: 4H structure,
PAC\
measurement at 150\textdegree C. BaMnO$_3$-1: 6H structure, PAC\ measurement at 400\textdegree C. 
BaMnO$_3$-21: sample with the 6H structure, X-ray measurement
at room temperature (RT). 
BaMnO$_3$-22: The same sample with X-Ray measurement at 900\textdegree C. BaMnO$_3$-3: 15R
sample, PAC
measurement at 400\textdegree C, X-ray measurement at RT. }\label{XRD}
\begin{tabular}{lcccccc}
\hline\hline
Sample     |& Phases  |& R$_P$ |& R$_{wp}$ |& R$_{exp}$ |& $a$ |& $c$           
  \\
\hline
SrMnO$_3$-1 &   4H &  12  &    18             &     3.15 &5.4526&9.0841        
\\
BaMnO$_3$-1&  6H  & 6   &    10                  & 0.52 &5.6277&13.9730         
   \\
BaMnO$_3$-21&  6H  & 39   &    51                  & 0.54 &5.6652&13.9678       
     \\
BaMnO$_3$-22         & 6H &  32  &     44                 &  
0.45&5.7162&14.0316            \\
BaMnO$_3$-3&15R& 3&4&0.65&5.6978&35.3354
        \\
\hline
\end{tabular}
\end{table}

\subsection{Magnetization}
\begin{figure}[tpb]
\begin{center}
\includegraphics[width=0.55\linewidth]{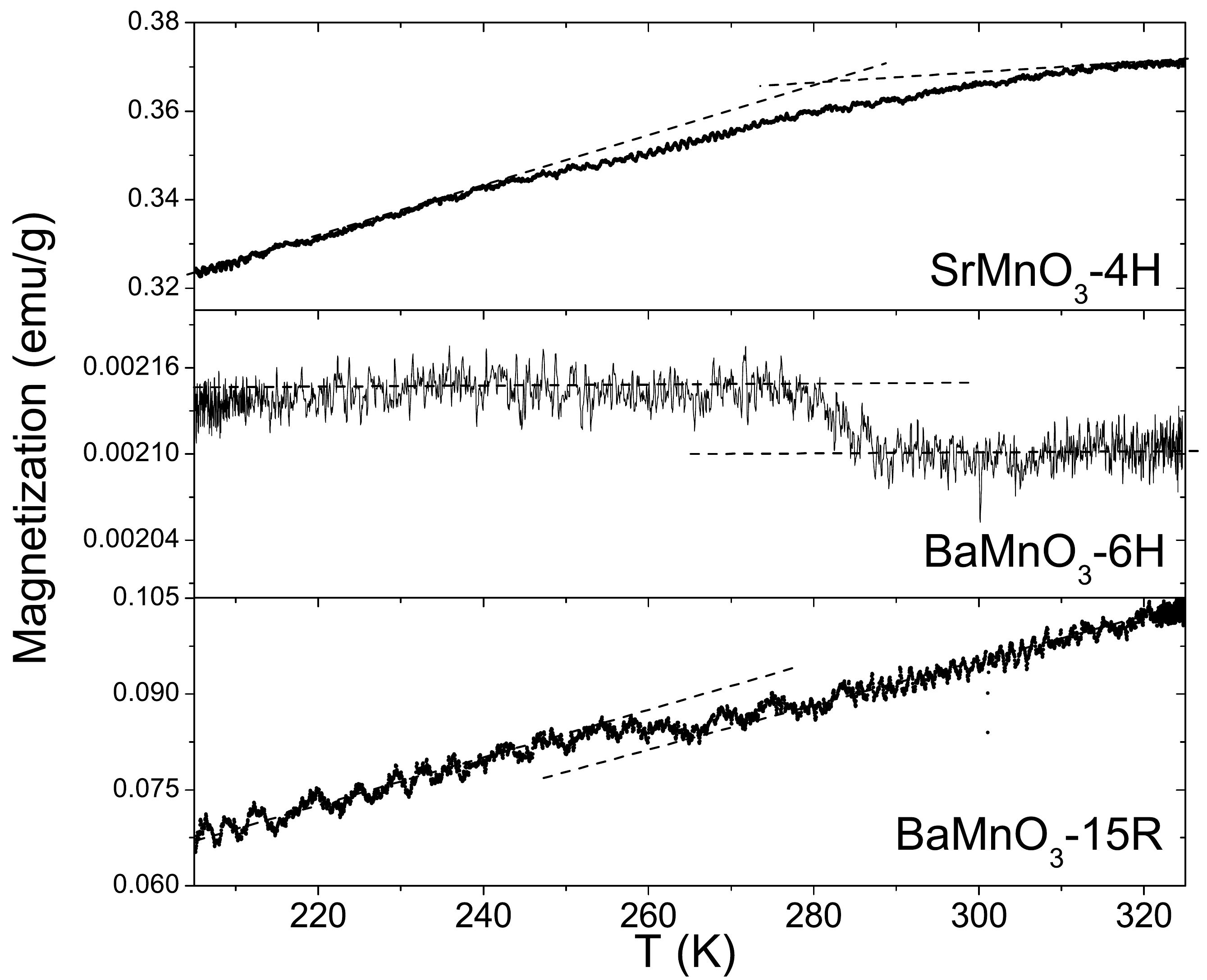}
\caption[]{Magnetization as a function of temperature 
near the magnetic anomalies for the samples:
\emph{Upper figure}: SrMnO$_3$-4H. \emph{Middle figure}: BaMnO$_3$-6H.
\emph{Lower figure}: BaMnO$_3$-15R. The lines highlight the crossover of the magnetization behavior in the transition region. }
\label{magnetic}
\end{center}
\end{figure}

Measurements of magnetization temperature dependence 
were also performed in the studied samples, one for each phase. 
A vibrating sample magnetometer was used, with magnetic 
field $B=5$ T for SrMnO$_3$-4H and BaMnO$_3$-15H, and $B=0.5$ T for
BaMnO$_3$-6H. The results are presented in figure~\ref{magnetic}. Small anomalies
are detected, signalling 
magnetic transitions, in SrMnO$_3$, at $280$ K ($7$ $^{\circ}$C) 
(consistent with previous measurements~\cite{Battle1988}), which are much sharper in the cases of BaMnO$_3$-6H, 
at $280$ K, and in BaMnO$_3$-15R, at $260$ K ($-16$ $^{\circ}$C) (consistent with~\cite{Adkin2007}). 

\subsection{PAC - SrMnO$_3$}
\begin{figure}[ht]
\subfigure{
\includegraphics[width=0.5\linewidth]{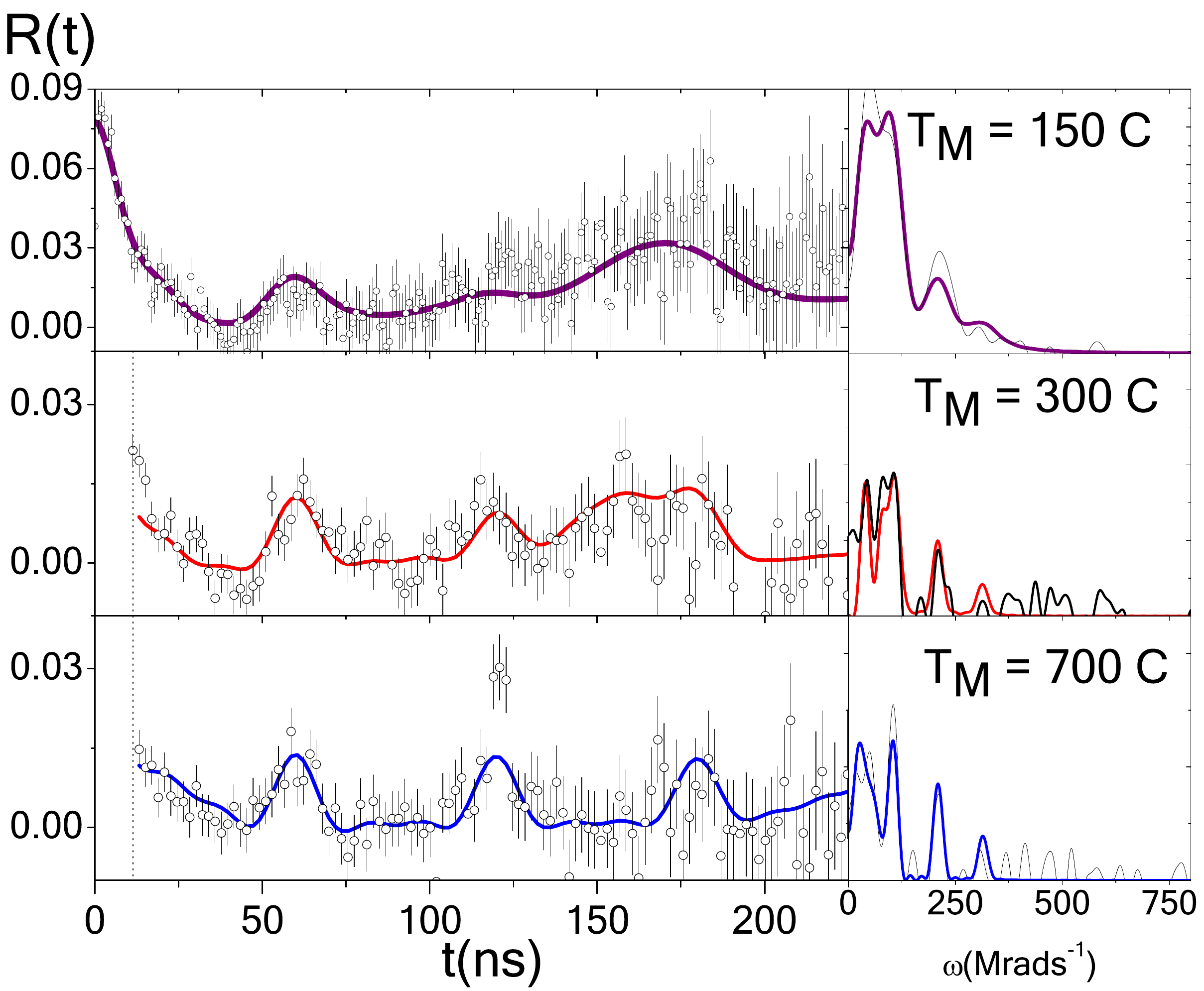}
\label{SMO_Vzzfit_simul}
}
\subfigure{
\includegraphics[width=0.5\linewidth]{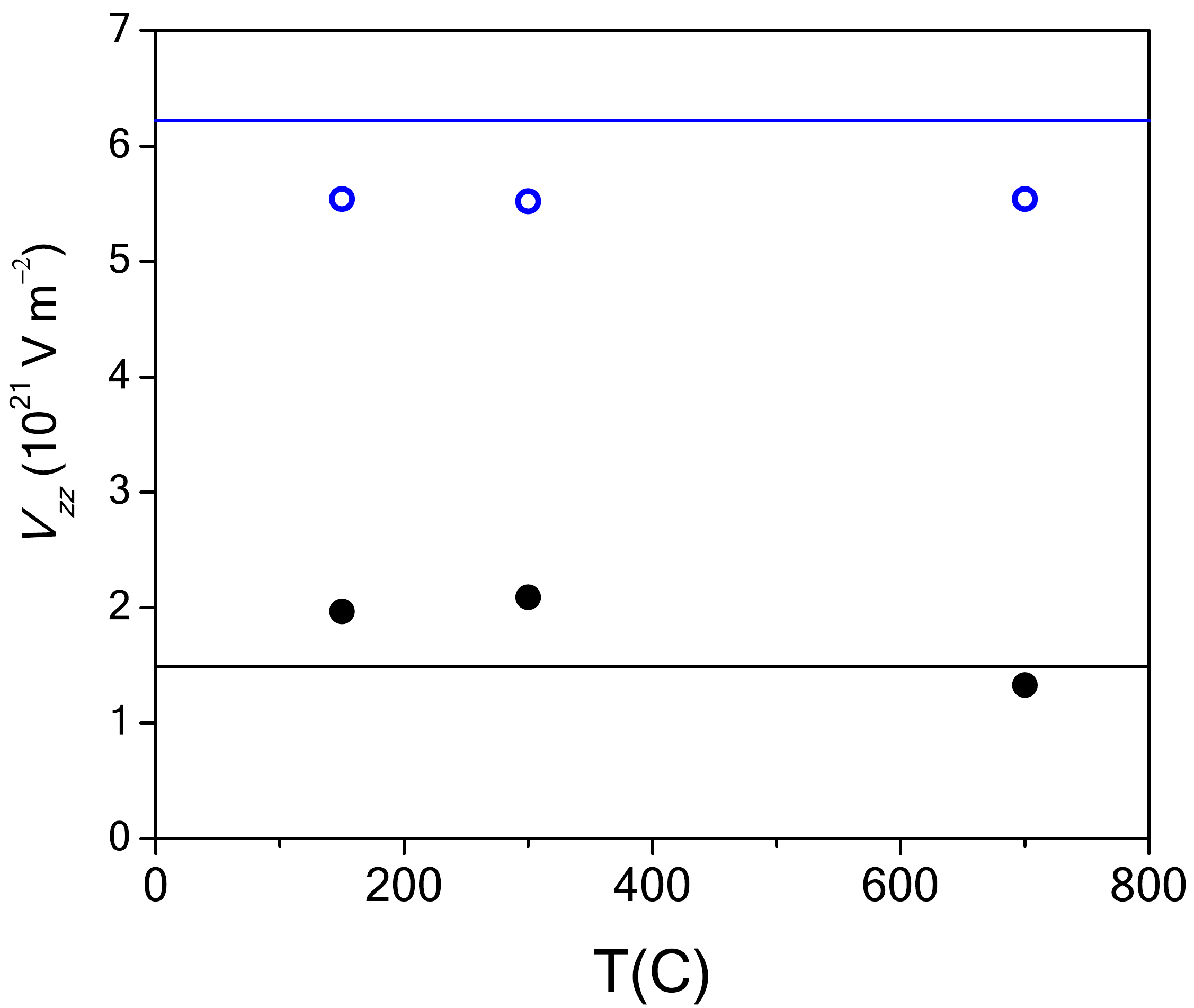}
\label{SMO_fit}
}
\caption[]{SrMnO$_3$-4H. \emph{Left figure}: the lines
represent 
the fits to the spectra of the perturbation
functions, at temperatures of 150 (up) , 300 (middle) and 700\textdegree C (down).
For all three spectra, the 
frequencies taken from the fits, shown in the adjacent Fourier transforms, are
approximately equal. 
\emph{Right figure}: The circles represent the experimental $V_{zz}$ obtained
with two fractions (open and closed circles). 
The horizontal lines represent the $V_{zz}$ obtained by DFT calculations
with highly diluted Cd probes substituting either one of the two
inequivalent sites in the structure.}
\label{SMO_fits}
\end{figure}

The SrMnO$_3$ PAC spectra for temperatures of 150, 300 and 700\textdegree C are shown in
figure~\ref{SMO_Vzzfit_simul}. 
The spectra can be reasonably fitted with two distinct local environments, 
fractions of 50 \%. $V_{zz}$ are shown for 50/50 fractions and asymmetry fixed
to 0 in figure~\ref{SMO_fit}. 
However, the 50\% values have a large estimated 20\% uncertainty, due to the statistical error and highly attenuated spectra. In this interval reasonable fits are still achieved.

Assuming the fractions fixed at 50/50 and asymmetry values fixed at zero
(also not crucial to the fit with these spectra), the higher, best defined value
of 
$V_{zz}$, and the lower $V_{zz}$ are found to be almost constant in the
temperature range (150-700\textdegree C).
This is consistent with the similar behavior in the CaMnO$_3$
system~\cite{Lopes2008}. 
The calculations of the two substitutions of Cd in the two Sr sites, shown
earlier,  give
two values of $V_{zz}$ in qualitative agreement,
as the values are rather close (figure~\ref{SMO_fit}). 
The small differences from the calculations in the higher 
value may be due to both temperature effects and the incomplete 
convergence of $V_{zz}$ with the size of the supercells.  
These results suggest that the Cd probe occupies substitutionally both Sr sites,
which is in agreement with the similar total energies predicted 
by the calculations.
The $V_{zz}$ of lower magnitude is almost constant as, far as we can tell, 
since given the limited time window the fit of a low EFG is not very accurate.


\subsection{PAC - BaMnO$_3$}

\begin{figure}[ht]\centering
\subfigure{
\label{BMO_PAC_spectra}\includegraphics[width=0.5\linewidth]{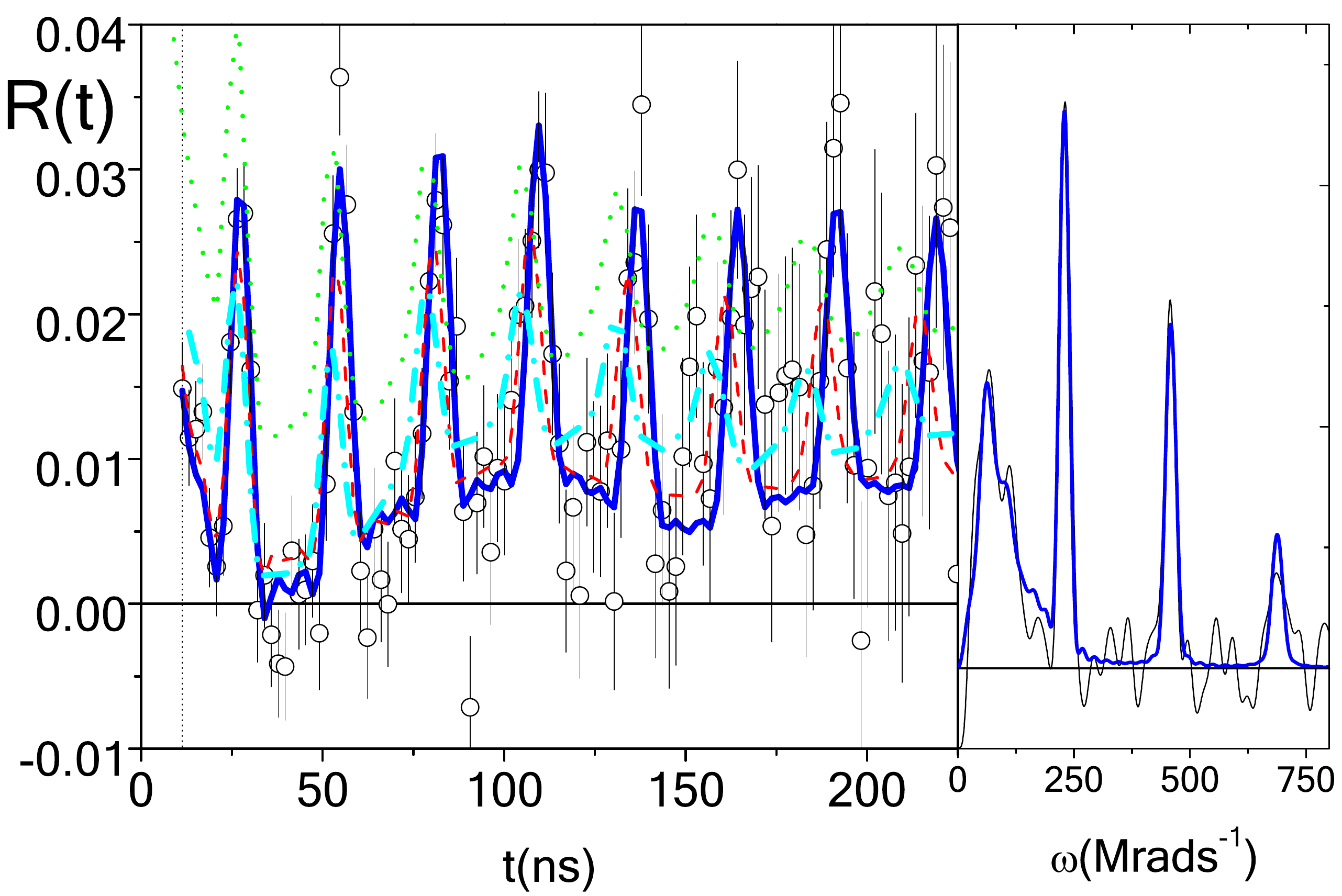}}
\subfigure{\label{BMO6HCdpf}\includegraphics[width=0.5\linewidth]{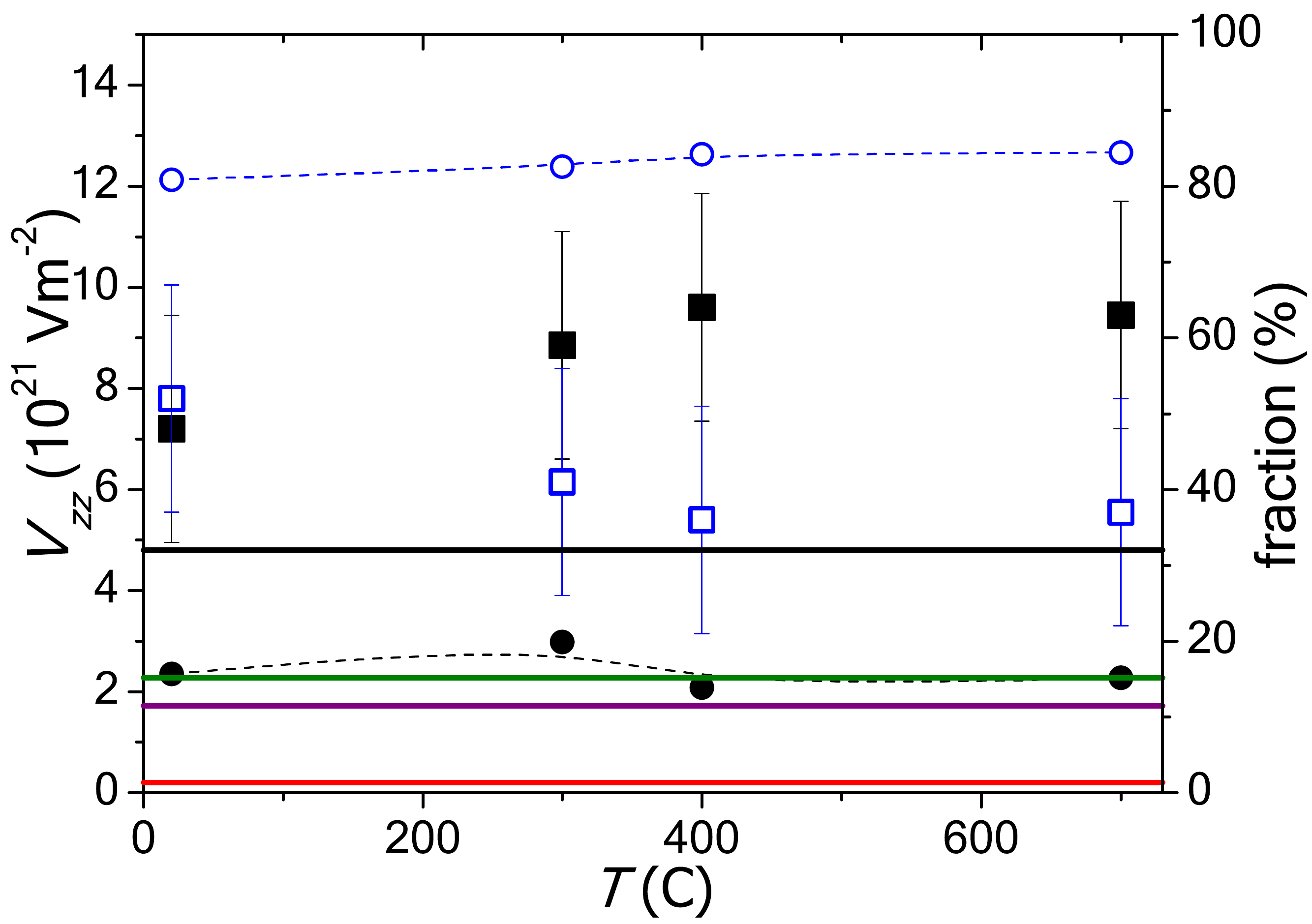}}
\caption{(a) Experimental perturbation spectra (dots with error
bars) for the BaMnO$_3$-6H sample at RT (left)
and a representative Fourier Transform (at RT). 
The lines in the left are fits to the R(t) spectra at different temperatures, 
showing quadrupole frequencies approximately constant: at RT (full line), 150\textdegree C
(dashed  line), 400\textdegree C(dotted  line), and 700\textdegree C (dash-dotted line). 
(b) The circles represent the experimental EFG of BaMnO$_3$-6H 
implanted with $^{111m}$Cd. There are two distinct environments measured, 
about $12$ and $2.5 \times 10^{21}$ V\,m$^{-2}$, differentiated by empty or full
symbols, respectively. The fraction of each environment is represented by the
squares, full or empty, according to the corresponding $V_{zz}$.
The four lines are the results of the calculations 
(disregarding the sign), considering a Cd probe
substitutional 
at each one of the four inequivalent Ba atoms in the 6H structure
(table~\ref{BMO6HCd}).}
\end{figure}

The spectra for the 6H-BaMnO$_3$ taken between 16 and 700\textdegree C do not present
appreciable frequency changes,
 and similarly the obtained EFG are constant, as shown in 
Figs.~\ref{BMO_PAC_spectra} and \ref{BMO6HCdpf}.
In this case the comparison with the first-principles calculations 
shows that one of the experimental frequencies measured is too
high ($\sim 12 \times 10^{21} $ V\,m$^{-2}$) to be 
accounted by any of the values
calculated with the 
Cd probe substitutional in a Ba site, even according to our 
previous (\textit{hhh}) BaMnO$_3$-2H supercell calculation 
which gives $V_{zz}\sim8 \times 10^{21} $ V\,m$^{-2}$.
Surprisingly  this is also the most well defined fraction of the spectrum. 
Its fraction is very large (40-50\%), of a unique local
environment so it is unlikely that it it comes from a different phase, 
as such a large fraction should be detected in the powder X-ray spectra, 
in spite of the difficulties to distinguish between the polymorphs.
Since the sample is a polycrystal, it may be due to probes located at or near
grain boundaries. We may also speculate that the corresponding 
Cd probes are favorably located
near vacancies or some other kind of defect, producing this high EFG. 
One test calculation, a
$2\times 2\times 1$ supercell with Cd substitutional at one of the sites and an
oxygen vacancy as a nearest neighbor produced, after atomic relaxations, a
relatively small $V_{zz} = 3.8 \times 10^{21} $ V\,m$^{-2}$. 
It is however possible that other sites for vacancies or vacancy-probe
separations may explain the present result. 

The lower EFG values of the other fractions may be interpreted as substitutional
Cd atoms at the Ba sites since its value is  close to the different values
calculated at the Cd atom substituting the inequivalent Ba sites, as can 
be seen by the lines in figure~\ref{BMO6HCdpf}. 
It is especially close to the Ba1 and Ba3 site 
calculations, which is consistent with the fact that 
these sites are predicted to favor Cd substitution 
over the others (table~\ref{BMO6HCd}). 

Note that the fractions have a high estimated average error, shown in the figure by error bars, so that an increase/decrease of the different environments with temperature cannot be established definitely, although an increase of the lower $V_{zz}$ fraction is likely. We further estimate a high accuracy for the higher $V_{zz}$ ($\pm 0.03$) while the lower $V_{zz}$ is less accurate ($\pm 0.3$).

\begin{figure}[ht]\centering
\subfigure{\label{BMO6HIn}\includegraphics[width=0.5\linewidth]{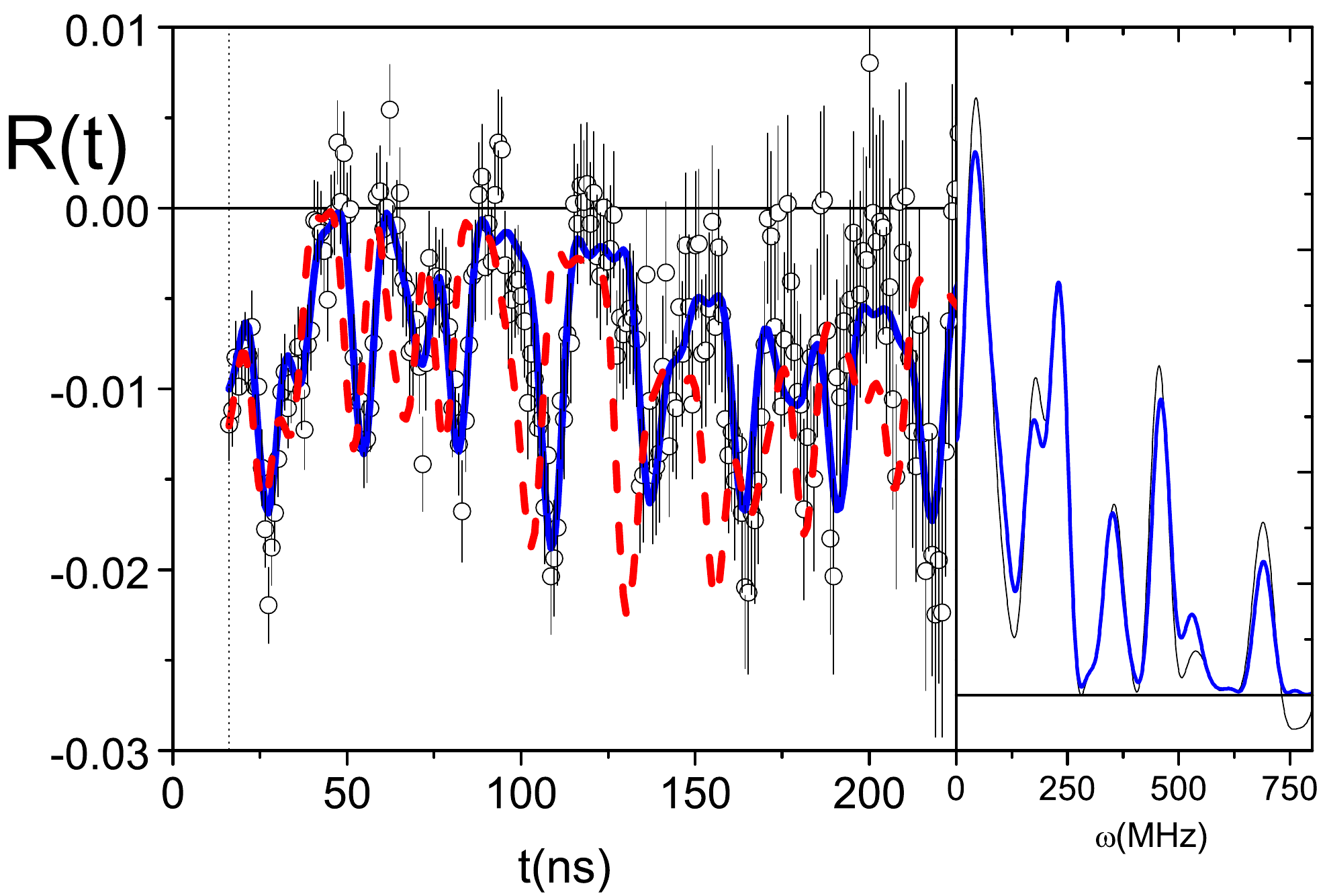}}
\subfigure{\label{BMO6HInpf}\includegraphics[width=0.5\linewidth]{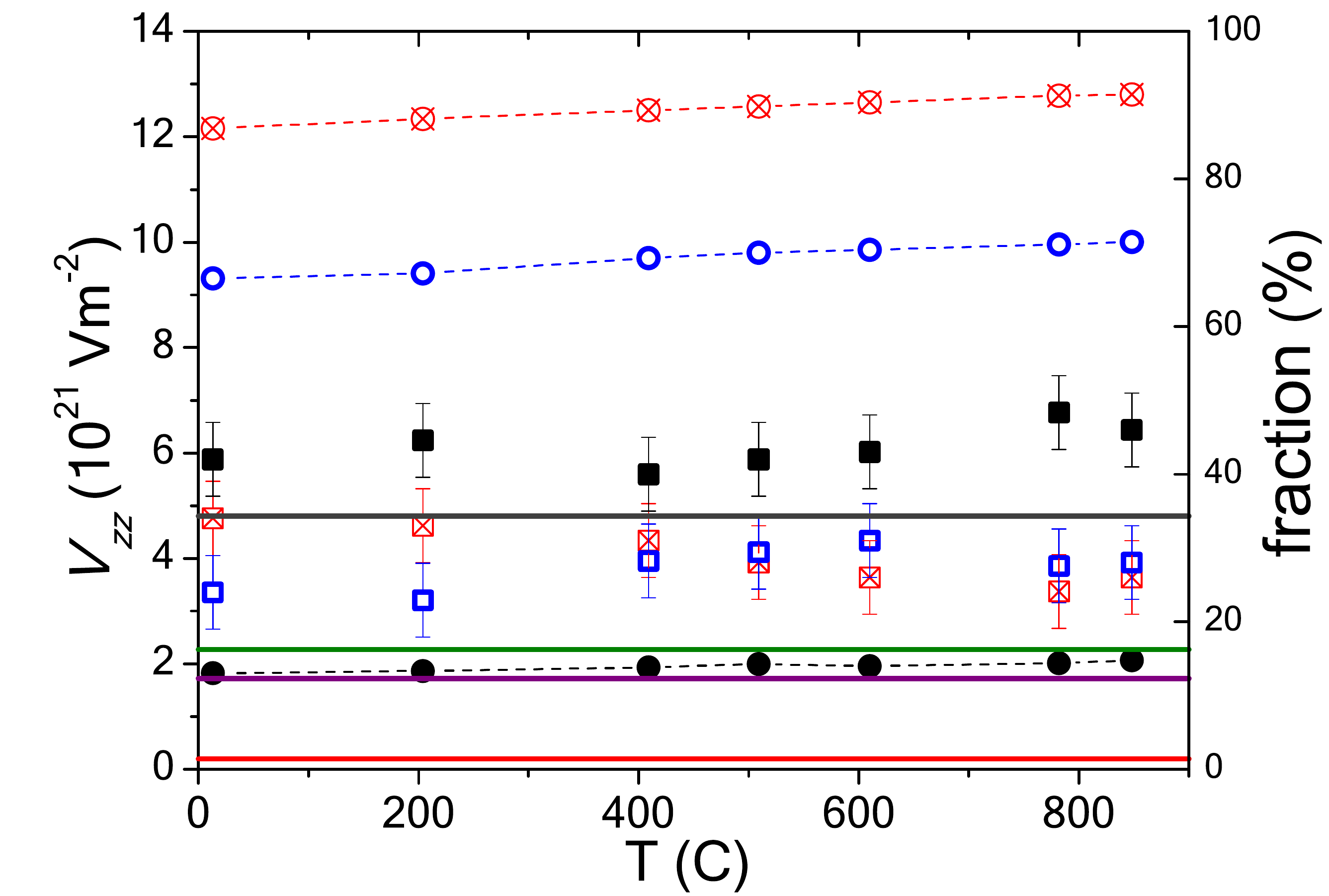}}
\caption{(a) Experimental perturbation spectra (left) and Fourier
Transforms
(Right), for the BaMnO$_3$-6H sample measured at RT(full line)
and 850\textdegree C
(dashed line), with the probe $^{111}$In/$^{111}$Cd. The Fourier transform plot
is for room
temperature. (b) The circles represent the 
experimental EFG of BaMnO$_3$-6H implanted
with
$^{111}$In. There are three distinct environments measured, about $12$, $9.5$
and $2 \times 10^{21}$ V\,m$^{-2}$, differentiated by the symbols (crossed,
empty and
full, respectively) in the plot. The fraction of each environment is represented
by the corresponding squares. The solid lines are the results of four
calculations (disregarding the sign), considering a Cd probe
substitutional at each one of the four inequivalent Ba atoms in the 6H
structure.}
\end{figure}

In this case, to get more information and confirm the previous measurements, 
we have also implanted the probe $^{111}$In/$^{111}$Cd on 6H-BaMnO$_3$ 
samples. 
The initial state of the $\gamma-\gamma$ cascade is different from the
$^{111m}$Cd 
and has different anisotropy coefficients, but the intermediate state of the
$\gamma$-$\gamma$ cascade is the same, 
and if the $^{111}$In probe goes to the same lattice sites as
$^{111m}$Cd when implanted and annealed, we 
should get the same quadrupole frequencies.  
The samples were measured with this probe  in the temperature range 16-850\textdegree C. 
The spectra are, as expected, almost constant on this range of temperatures (13,
16, 206, 409, 510, 611, 782 and 858\textdegree C). 
R(t) representative functions and its Fourier transforms are shown in
figure~\ref{BMO6HIn}. 
Unlike the Cd probe case, the fit has to be done not with two, but three
fractions.  There are two fractions in agreement with the $^{111m}$Cd case and
there is a third fraction, 
also with $V_{zz}\sim 9 \times 10^{21} $ V\,m$^{-2}$, higher than with the
calculations. This value is closer to 
the $V_{zz}\sim8 \times 10^{21} $ V\,m$^{-2}$ calculated 
for the (\textit{hhh}) environment, therefore the Ba2 site 
substitution might be 
an additional effect caused by the implantation of $^{111}$In,
which in this case could not be
removed with thermal annealing (which is the same as in $^{111m}$Cd
implantations). 
There is a slight increase of the frequencies of the  higher $V_{zz}$ with
increasing temperature, as in the $^{111m}$Cd probe case. 
Therefore, in this case a fraction of the probes ($\sim 43\%$) must still be
substitutional at the Ba sites 1 or 3, with the 
lower $V_{zz}$, probably another 
fraction at the the Ba 2 site, and for the
other environment we still cannot assign a specific site. The three environments have average values 26.2\%, 27.3\%, and 43.7\%.  The value of the fractions of each environment has an estimated error bar of around 5\%.
The results obtained at different temperatures are consistent with reasonably stable fractions of the three reported environments. The corresponding error bar in $V_{zz}$ is $0.1 \times 10^{21}$\ V\,m$^{-2}$.

\begin{figure}[ht]\centering
\includegraphics[width=0.5\linewidth]{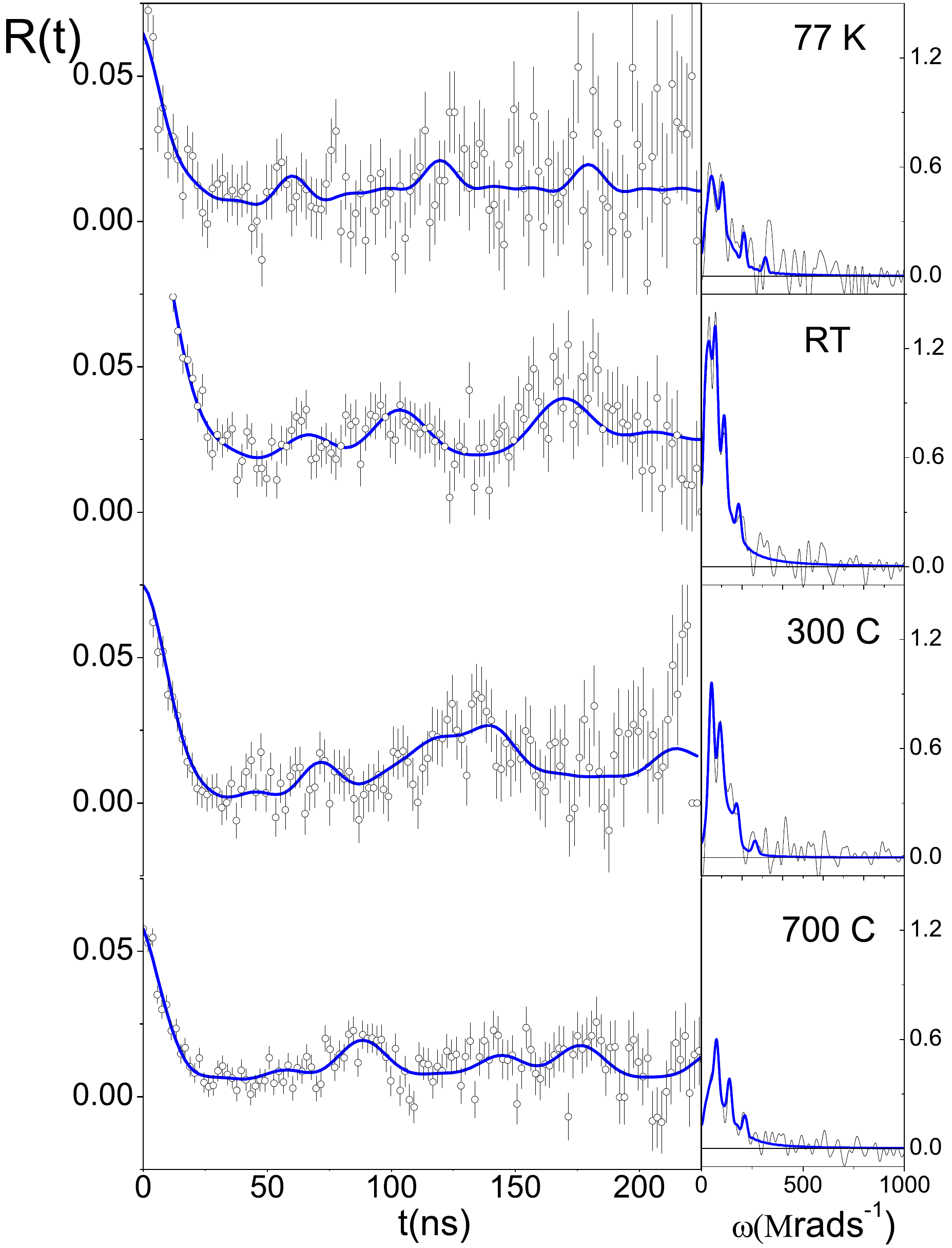}
\caption{Experimental perturbation spectra (left) and Fourier
Transforms (Right)
for the 15-R sample,
show that the quadrupole frequencies are low at all temperatures
measured.}\label{15Rall}
\end{figure}

Figure~\ref{15Rall} shows measurements in the 15-R polymorph samples. 
The perturbation functions are quite attenuated and low frequencies are 
measured at all temperatures 77 K, RT, 300\textdegree C and 700\textdegree C.

At room temperature and above, two different EFG environments clearly give a
much better fit than just one EFG.
At 77 K, due to the large statistical error, this is not so clear, 
but in order to be consistent with the RT measurement, two EFGs
were used in all fits.
The fractions of the two environments are approximately  $84\%$ and $16\%$ in
all the fits. Again, for the low temperature  there was a large range allowed
due to the larger statistical error, but the fits at all temperatures used
starting values of $f_1=84\%$ and
$f_2=16\%$ (values obtained by fitting the RT measurement), and the final values
do not change significantly in any case.
 The experimental EFG parameters for both fractions are shown in table
\ref{PAC15REFG}. 

The $V_{zz}$ parameters are interspersed values on the range $1.89$ to $3$ for
fraction f$_1$ and $3.1$ to $5.6\times 10^{21}$ V\,m$^{-2}$ for fraction f$_2$. 
These small variations may exist due to structural variations with temperature
or dynamic effects, but these values are close to the calculated values at Cd 
in the 6H polymorph, indicating also
substutional Cd at the Ba sites in this case. 

\begin{table}[ht]\centering
\caption{$V_{zz}$ ($10^{21}$ V\,m$^{-2}$) taken from the
PAC measurements of BaMnO$_3$,
in the rhombohedral 15R polymorph, at four different temperatures.
At 30\textdegree C the spectrum was fitted with two environments,
in that case the $V_{zz}$ of the left column is fitted with a fraction of
$84\%$,
and the right column $V_{zz}$ corresponds to $16\%$.}\label{PAC15REFG}

\begin{tabular}{lrr}
\hline\hline
T                 & V$_{zz}$ of f$_1$    & $V_{zz}$ of f$_2$\\
\hline
77 K &               2.58           &   5.55   \\
30 C &               1.89           &   3.12   \\
300 C  &             2.62           &   4.66   \\
700 C  &             3.00           &   3.72   \\
\end{tabular}
\end{table}

\section{S\lowercase{r}$_{1/2}$B\lowercase{a}$_{1/2}$M\lowercase{n}O$_3$}
\label{SBMO}
In this section we study the EFGs of the constituent 
elements in the mixed perovskite Sr$_{1/2}$Ba$_{1/2}$MnO$_3$ 
and its variation with electric polarization. 
In this case we used the projector augmented wave (PAW) 
method~\cite{Blochl1994}, as implemented in \textsc{vasp}~\cite{Kresse1996}.
Following the previous DFT study of this case~\cite{Giovannetti2012}, 
we minimize the total energy relaxing the polar displacements 
of Mn and O atoms, with the experimental $a=3.85\,$\AA{}, 
and we also consider both $c/a = 1.01$ 
and $c/a = 1.005$, which are representative experimental values of the 
paramagnetic and G-AF phases, respectively 
(see Sakai et al.~\cite{Sakai2011}, figure~3). 
The same GGA+U approach is used 
with $U=4.5$ eV and $J=1.0$ eV for the Mn $d$ electrons,. 
The G-type antiferromagnetic order is considered, and we 
assume the ordered structure for the Ba/Sr ions 
in the same alternating fashion. 

After relaxation of the ionic position to forces less than $0.01$ eV/\AA{}
 we obtain a spontaneous polarization 
 of $10\,\mu$C\,cm$^{-2}$, with $c/a=1.01$, which is smaller  
than the value obtained by Giovannetti et.\ al.~\cite{Giovannetti2012}, 
and consistently also Mn-O-Mn angles closer to $180^{\circ}$.
This is probably due to different PAW potentials used,  
which also result in different ionic and electronic contributions. 
Table~\ref{polSBMO} summarizes the results. 
The calculation for $c/a=1.005$ follows the 
expected tendency to supress ferroelectricity, mainly with 
a change of electronic polarization, 
consistent with~\cite{Giovannetti2012}.
 
 \begin{table}[ht]\centering
\caption{Ionic, electronic and total 
electronic polarization ($\mu$C\ cm$^{-2}$), 
and Mn-O-Mn angle $\alpha$ ($^{\circ}$) 
for $c/a=1.01$ and $c/a=1.005$. }\label{polSBMO}
\begin{tabular}{lrrrr}
\hline\hline
 $c/a$   & $P_{ion}$ &  $P_{ele}$ & $P_{tot}$ & $\alpha$\\
\hline
 $1.01$  & 2.02 & -12.09 & -10.06 & 177.89 \\
 $1.005$ & 2.39 &  -7.64 &  -5.25 & 178.82 \\
 \end{tabular}
\end{table}

\begin{figure}[ht]\centering
\includegraphics[width=0.5\linewidth]{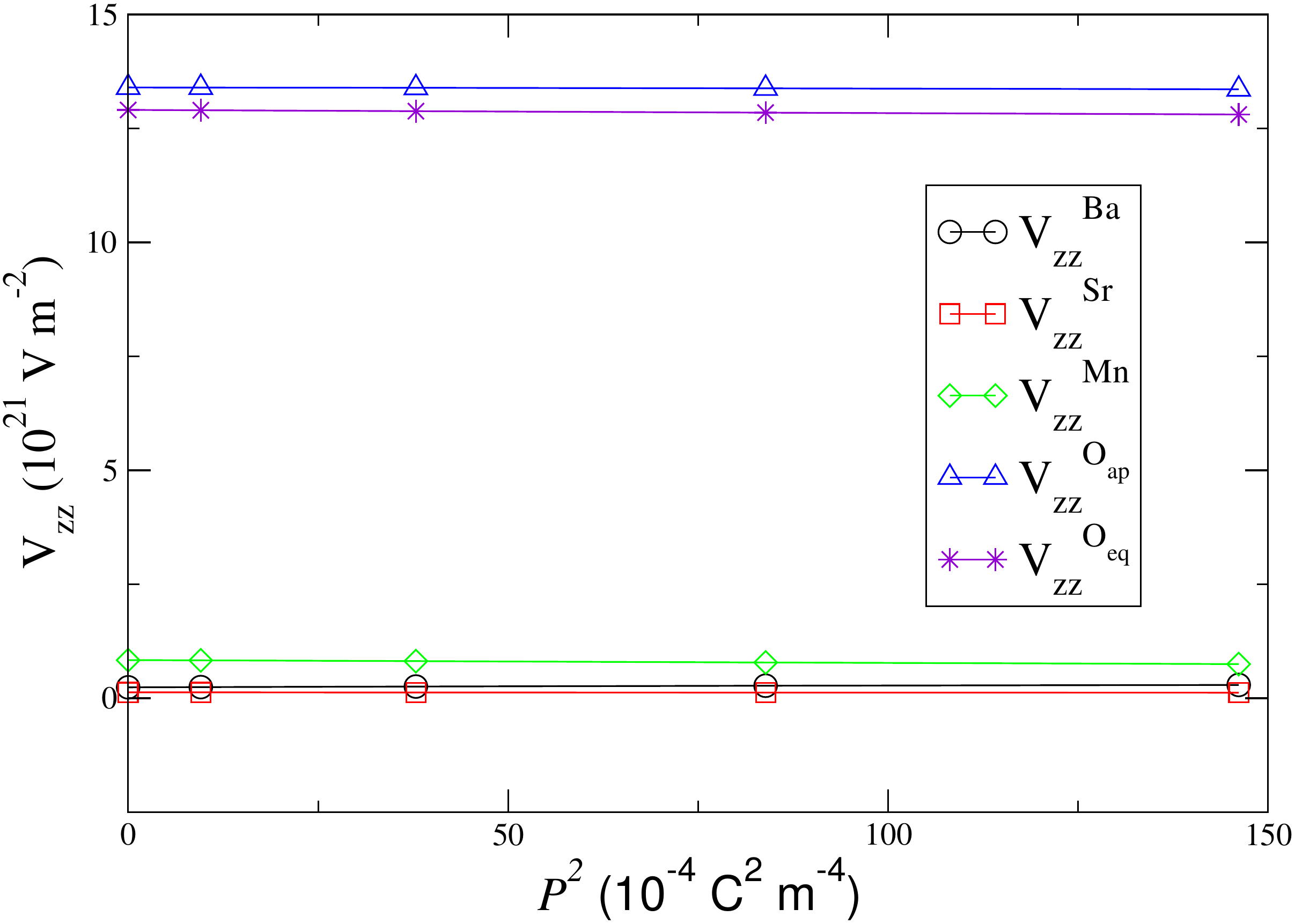}
\caption{$V_{zz}$ as a function of the polarization squared for 
the inequivalent atoms in the 
perovskite Sr$_{0.5}$Ba$_{0.5}$MnO$_3$.}\label{VzzP2SBMO}
\end{figure}

Using the $c/a=1.01$ structure as a reference we considered 
a set of structures in the adiabatic path connecting this structure 
to the centrosymmetric ($P4/mmm$) structure, 
by changing the amplitude of the polar distortion. 
In this way we can check what is the sensitivity of the EFG with distortion,  
and the possibility of probing polarization differences via EFG variations.
The polarization varies linearly with the distortion, 
as is the case for other displacive ferroelectrics such as BaTiO$_3$.
The variation of EFG with the distortion (polarization) 
is quadratic, also similar to BaTiO$_3$~\cite{Gonçalves2012}.
However, contrary to that case, the EFG is 
almost insensitive to the polarization changes. 
figure~\ref{VzzP2SBMO} presents the variation of 
$V_{zz}$ against $P^2$ for the inequivalent atoms of the structure. 
The values appear constant in the whole range studied. 
The Ba/Sr/Mn atoms have always values close to $0$, which would be the case 
in the ideal cubic structure. 
The O site would have a high $V_{zz}$ in the cubic structure, close 
to the ones calculated here for all distortions. 
In fact they have regular variations discernable by the 
calculations, as shown in table~\ref{VzzP2SBMOtab}, 
 but still very small. $V_{zz}$ at 
the Ba atoms increases with distortion while it decreases for the other atoms. 
However, even this small variation appears only 
for artificially large polarization variations which are unlikely to be 
found. 

One fact helps to understand 
this small variation of the EFG, in 
distinction to BaTiO$_3$~\cite{Gonçalves2012}.
Here, the Mn and O atoms are displaced $\sim0.02$ \AA{}, 
whereas in BaTiO$_3$ the 
displacements are $\sim0.8$, $\sim0.5$ and $\sim0.2$ \AA{}, 
and already in that case the 
EFG did not change much ($\sim1\times 10^{21} $V\,m$^{-2}$ on 
average, in the complete distortion range). 
(However, note that the EFG is generally a function 
of the electron density, and its variation cannot be attributed 
to displacements only. 
In other cases, a change of magnetic order while keeping  
the same structure is enough to change the EFG at some atoms, such as 
hexagonal rare-earth manganites and YMn$_2$O$_5$~\cite{Note1}.)
In the present case, the small changes, though essential 
to stabilize the considerable polarization, are not enough to produce 
appreciable changes in the EFG. By considering also 
the $c/a=1.005$ case (table~\ref{VzzP2SBMOtab}, last row), 
with atomic relaxation, one can also check approximately the effect of the 
structural variation which is seen from the 
paramagnetic ($c/a\sim1.01$) to G-AF ($c/a\sim1.005$) order. 
Now the effect is larger, but still small, and may not 
be easily distinguised in an experience, 
where factors such as the statistical data acquisition error may 
decrease the accuracy of the measurements. 
Therefore, we conclude that the EFG at every site from the static 
contribution is approximately constant at 
the ferroelectric phase. 
However, this also brings an opportunity, since in this case the 
dynamic effects should 
dominate the EFG variation. This could be useful to analyze the 
phase transition, due to the increasing fluctuations near $T_C$, where the EFG 
is expected to present an abrupt variation similar to the divergence of 
electric susceptibility~\cite{Lopes2008,Yeshurun1978}.

\begin{table}[ht]\centering
\caption{$V_{zz}$ ($10^{21}$ V\,m$^{-2}$) for the different atoms 
(Ba, Sr, Mn, apical and equatorial O) taken from the
calculations for different values of the polar distortion (with $c/a=1.01$), 
where $\lambda = 1$ is the distortion found by the force minimization procedure. 
Last row: calculation for equilibrium 
structure with $c/a=1.005$. }\label{VzzP2SBMOtab}
\begin{tabular}{lrrrrrr}
\hline\hline
 $\lambda$ & $V_{zz}^{Ba}$ & $V_{zz}^{Sr}$ & $V_{zz}^{Mn}$ 
 & $V_{zz}^{O_{ap}}$ & $V_{zz}^{O_{eq}}$ \\
\hline
0    & 0.241 & 0.129 & 0.837 & 13.400 & 12.907\\
0.25 & 0.245 & 0.129 & 0.831 & 13.398 & 12.900\\
0.5  & 0.258 & 0.128 & 0.814 & 13.392 & 12.880\\
0.75 & 0.279 & 0.126 & 0.786 & 13.379 & 12.848\\
1    & 0.292 & 0.123 & 0.745 & 13.359 & 12.807\\
$c/a=1.005$ & 0.136 & 0.064 & 0.403 & 13.169 & 12.906 \\
\end{tabular}
\end{table}

\section{Summary}
We present a hyperfine local probing of BaMnO$_3$-6H, -15R, 
and SrMnO$_3$-4H manganites. 
The values of the EFG were measured with the $^{111m}$Cd and $^{111}$In probes
using the PAC
method.  It is found that $V_{zz}$ is roughly constant in a
range of temperatures 20-700\textdegree C, for both BaMnO$_3$-6H and SrMnO$_3$.
First-principles calculations of the EFG with the accurate (L)APW+lo method are 
reported. The calculated EFG at all atoms is shown, with SrMnO$_3$ in the low
temperature structure, and various polytype structures of BaMnO$_3$. The
values are largely insensitive to the magnetic order or approximation
used, and the relaxed structures give consistent results the experimental
structures.  For direct comparison with the experiments, we performed
calculations with supercells and introducing a Cd impurity substitutional at the
alkaline-earth sites. In the case of SrMnO$_3$ there is a good qualitative
agreement, which shows that the Cd probe occupies the Sr sites. For the case of
BaMnO$_3$-6H, although the calculated values are of the order of one of the
measured $V_{zz}$ values,
other high values remain incompatible with the divalent alkaline-earth
substitution. For BaMnO$_3$-15R, the EFG has values of the order expected if Cd
occupies the Ba sites.

Finally, we analyze the EFGs in the multiferroic perovskite 
manganite Sr$_{1/2}$Ba$_{1/2}$MnO$_{3}$, and find that 
they should be approximately constant for a wide range of structures.

This work is supported by the program COMPETE/FFEDER and FCT under projects 
PTDC/FIS/105416/2008, CERN/FP/123585/2011 and CICECO – (PEst-C/CTM/LA0011/2011 
and PEst-C/CTM/LA0011/2013). Further 
support from the ISOLDE collaboration, with 
approved project IS487, European Union Seventh Framework 
through ENSAR (contract no.\ 262010) and the BMBF German research program under 
contracts 05KK7TS2 and 05K10TS2 is acknowledged.
JNG acknowledges his postdoctoral grant from FCT (SFRH/BPD/82059/2011). 
We thank H.\ Haas for useful discussions. 


\end{document}